\documentclass[letterpaper,titlepage,11pt]{article}
\usepackage[colorlinks=true,urlcolor=blue,citecolor=magenta]{hyperref}
\usepackage{amssymb,amsmath,amsfonts}
\usepackage{epsfig}
\usepackage{epsfig}
\usepackage{graphicx}
\usepackage{epstopdf}
\usepackage{caption}
\usepackage{subcaption}
\usepackage{amscd}
\usepackage{amsthm}
\usepackage{latexsym}
\usepackage{amsbsy}
\usepackage{bbm}
\usepackage[english]{babel}
\usepackage{psfrag}
\usepackage{tabularx}

\def\be{\begin{equation}}
\def\ee{\end{equation}}
\def\bea{\begin{eqnarray}}
\def\eea{\end{eqnarray}}

\allowdisplaybreaks

\def\clock{{\count0=\time
           \divide\count0 60
           \ifnum\count0<10 0\fi\the\count0
           \multiply\count0 -60 \advance\count0 \time
           :\ifnum\count0<10 0\fi \the\count0
         }}
\newcommand{\timestamp}{{\small\vbox{\hbox{\tt\jobname.tex}
\hbox{\the\day/\the\month/\the\year, \clock}}}}

\def\sp{\;\;\;,\;\;\;}

\numberwithin{equation}{section}

\begin{document}

\begin{titlepage}
\rightline{\vbox{   \phantom{ghost} }}

 \vskip 1.8 cm

\centerline{\LARGE \bf
Lifshitz Hydrodynamics from Lifshitz}
\vskip .3cm
\centerline{\LARGE \bf  Black Branes with Linear Momentum}

\vskip 1.5cm

\centerline{\large {{\bf Jelle Hartong$^1$, Niels A. Obers$^2$, Marco Sanchioni$^2$}}}

\vskip 1.0cm

\begin{center}

\sl $^1$ Physique Th\'eorique et Math\'ematique and International Solvay Institutes,\\
Universit\'e Libre de Bruxelles, C.P. 231, 1050 Brussels, Belgium.\\
\sl $^2$ The Niels Bohr Institute, Copenhagen University,\\
\sl  Blegdamsvej 17, DK-2100 Copenhagen \O , Denmark.
\vskip 0.4cm

\end{center}

\vskip 1.3cm \centerline{\bf Abstract} \vskip 0.2cm \noindent

We construct a new class of 4-dimensional $z=2$ Lifshitz black branes that have a nonzero linear momentum. These are solutions of an Einstein--Proca-dilaton model that can be obtained by Scherk--Schwarz circle reduction of AdS$_5$ gravity coupled to a free real scalar field. The boundary of a bulk Lifshitz space-time is a Newton--Cartan geometry. We show that the fluid dual to the moving Lifshitz black brane leads to a novel form of Lifshitz hydrodynamics on a Newton--Cartan space-time. Since the linear momentum of the black brane cannot be obtained by a boost transformation the velocity of the fluid or rather, by boundary rotational invariance, its magnitude plays the role of a chemical potential. The conjugate dual variable is mass density. The Lifshitz perfect fluid can be thought of as arising from a Schr\"odinger perfect fluid with broken particle number symmetry.

\end{titlepage}

\newcommand{\red}[1]{{\color{red} #1 \color{black}}}
\newcommand{\blue}[1]{{\color{blue} #1 \color{black}}}
\newcommand{\green}[1]{{\color{green} #1 \color{black}}}
\newcommand{\yellow}[1]{{\color{yellow} #1 \color{black}}}

\newcommand{\EK}[1]{{\red{EK: #1}}} 
\newcommand{\JH}[1]{{\blue{JH: #1}}} 
\newcommand{\NO}[1]{{\green{NO: #1}}} 
\newcommand{\JdB}[1]{{\yellow{JdB: #1}}} 


\tableofcontents

\section{Introduction}

Hydrodynamics is a powerful tool to describe the long wavelength physics of quantum field theories at finite temperature. 
Remarkably, holography provides a dual realization of such effective descriptions, typically at strong coupling in the field theory, 
in terms of the long wavelength dynamics of black holes. The seminal example of this has been the calculation 
of the viscosity to entropy ratio \cite{Policastro:2001yc,Kovtun:2004de} (see \cite{Son:2007vk}  for a review)  of strongly coupled plasmas using black holes 
via the AdS/CFT correspondence. This deep relation between fluid dynamics and gravity has led to the fluid/gravity correspondence \cite{Bhattacharyya:2008jc,Baier:2007ix} (see \cite{Rangamani:2009xk} for a review) which has sparked numerous novel insights on both sides
of the duality.%
\footnote{For asymptotically flat black branes the blackfold approach \cite{Emparan:2009cs,Emparan:2009at,Camps:2010br}
also gives a relation between the long wavelength dynamics of black branes and fluids that live on
 dynamically embedded surfaces. Applied to D3-branes
this has been shown to encapsulate AdS fluid/gravity \cite{Emparan:2013ila}.} 

These developments initially focused on the dual gravitational formulation of relativistic hydrodynamics, since in the standard AdS holography the dual
field theories are relativistic. Motivated by applying holography in a wider setting, in particular strongly coupled non-relativistic field theories, 
more general holographic bulk theories with anisotropic scaling between time and space, characterized by the dynamical exponent $z$, 
 have been introduced  \cite{Son:2008ye,Balasubramanian:2008dm,Kachru:2008yh,Taylor:2008tg}.
 These include  include Schr\"odinger and Lifshitz space-times, and in the latter case%
\footnote{See \cite{Taylor:2015glc} for a review on Lifshitz holography.}
 there are different bulk realizations (e.g. Einstein--Maxwell-dilaton  (EMD) and Einstein--Proca-dilaton (EPD) models) which have distinct physical features. 
 All these holographic models serve to describe different types of non-relativistic field theories, where in the former there is  Galilean boost symmetry, while in the latter there is a broken boost symmetry. For possible applications to condensed matter systems, but also to further our understanding of holography
in non-AdS setups, it is thus a natural question to generalize the fluid/gravity correspondence to these different classes of non-relativistic field theories. 
This may also serve as a step towards a more general classification of such field theories.

For theories with Schr\"odinger symmetries a corresponding version of (conformal) non-relativistic fluid/gravity 
correspondence was developed in \cite{Herzog:2008wg,Maldacena:2008wh,Adams:2008wt,Rangamani:2008gi}. 
Certain realizations of Lifshitz hydrodynamics and their holographic description have subsequently been considered in 
\cite{Hoyos:2013eza,Hoyos:2013qna,Hoyos:2013cba}.  Another class of Lifshitz theories and their hydrodynamics
was holographically studied in the context of the Einstein-Maxwell-dilaton (EMD) model  \cite{Kiritsis:2015doa}.
In these theories there is an extra bulk $U(1)$ symmetry and since the dilaton runs logarithmically close to the boundary, there is
a new scaling exponent on top of the dynamical exponent $z$. 

In this paper we will focus on yet another class of Lifshitz theories, namely those that have Lifshitz symmetry
in the bulk and Schr\"odinger symmetry with broken particle number on the boundary. 
For a large class of EPD models, it was shown in   \cite{Christensen:2013lma,Christensen:2013rfa,Hartong:2014oma}
that holography in such bulk theories is dual to non-relativistic field theories of this type, coupled to a background torsional Newton-Cartan geometry
that is induced on the boundary. 
Our aim is therefore to find a gravitational dual realization of the hydrodynamics, or rather the perfect fluid limit, of this class of non-relativistic theories by constructing  appropriate Lifshitz black branes.  

A classification of the different versions of Lifshitz hydrodynamics will be given in the upcoming work \cite{Hartong} using a field theory perspective. The novel version of Lifshitz hydrodynamics that we find in this paper is a holographic realization of one particular class, which will
be also discussed in  \cite{Hartong} with field theory examples.%
\footnote{See also \cite{Hartong:2015wxa} for a discussion on field theories coupled
to torsional Newton-Cartan geometry with broken Schr\"odinger symmetries.} 
We will find in this paper that this version of Lifshitz hydrodynamics requires the construction of a new class of four-dimensional
$z=2$ Lifshitz black branes that have a non-zero linear momentum. 
While, as mentioned above, a large class of general $z$ Lifshitz space-times can be constructed in the EPD model, we restrict for technical
reasons to a particular EPD model with $z=2$ solutions, that can be obtained by Scherk--Schwarz circle reduction of AdS$_5$ gravity coupled to a free real scalar field. 

Our new $z=2$ Lifshitz black brane solutions exhibit the following features:
\begin{itemize}
\item The linear momentum of the black brane cannot be obtained by a boost transformation, and hence
 this class of solutions is physically distinct from unboosted solutions.  
\item The (squared) magnitude of the boost velocity plays the role of a chemical potential dual to the mass density. 
Consequently, the mass density occurs asymptotically as an extra parameter on top of the energy, even when the velocity is zero. 
\item The black brane configurations describe a new class of Lifshitz perfect fluids that are obtained by breaking particle number symmetry in
Schr\"odinger perfect fluids. 
\end{itemize}
In further detail, the thermodynamics for these Lifshitz black branes can be summarized by 
\begin{eqnarray}
\label{eq:intro}
\mathcal{E}+P & = & Ts+\frac{1}{2}\rho V^2\,,\\
\label{eq:intro2}
\delta\mathcal{E} & = & T\delta s+\frac{1}{2}V^2\delta\rho\,,
\end{eqnarray}
where $\mathcal{E}$ is the energy density, $P$ the pressure with equation of state $P=\mathcal{E}$ (2 spatial dimensions with $z=2$), $T$ temperature, $s$ entropy density, $\rho$ mass density and $V^2=V^iV^i$ with $V^i$ the fluid velocity.

\subsubsection*{Outline}

The outline of the paper is as follows. In section \ref{sec:bulktheory} we introduce the Einstein--Proca-dilaton (EPD) theory which consists of Einstein gravity coupled to a massive vector and a dilaton with arbitrary dilaton-dependent couplings between the massive vector and the dilaton. In section \ref{subsec:Lifsol} we show that this model admits, under some mild restrictions on the dilaton-dependent coupling functions, Lifshitz solutions for any value of the dynamical exponent $z$. Important for the rest of this work we show in section \ref{subsec:dimred} that there is one EPD model that can uplifted to a 5-dimensional AdS gravity theory coupled to a free real scalar field. This specific model will be referred to as the upliftable model and it admits $z=2$ Lifshitz solutions.

For the higher-dimensional AdS theory it is known how to perform holographic renormalization and by reducing the result to one dimension lower we can obtain the relevant counterterms and near-boundary asymptotic expansions. This reduction is of the Schwarz--Schwarz type meaning that the 5-dimensional scalar field is required to come back to itself up to a shift (which is a global symmetry) upon going around the compact 5th dimension. In appendix \ref{app:dimred} we provide the details of the holographic renormalization before and after the dimensional reduction. In section \ref{subsec:consistency} we give a proof that the reduction is consistent.

The reduction in the bulk is everywhere along a spacelike circle however on the boundary (due to conformal rescaling) the circle is null. Hence from a boundary perspective we are dealing with a null reduction. It is well known that null reductions of Lorentzian geometries, in this case the boundary of the asymptotically AdS$_5$ space-time, lead to Newton--Cartan geometries. The details of this null reduction for both the metric and energy-momentum tensor are given in appendix \ref{app:TNC}.

Section \ref{sec:blackbranesmom} is concerned with the construction and properties of Lifshitz black branes with linear momentum. We start with the ansatz in section \ref{subsec:ansatz} where we also show that the effective action, that reproduces the equations of motion of the EPD model in which our ansatz has been substituted, possesses two scale symmetries. This leads to two Noether charges or first integrals of motion that are constant along the holographic coordinate. The following sections \ref{subsec:asympsol}--\ref{subsec:interpolating} study the properties of the solution near the boundary and near the horizon. In the last two subsections \ref{subsec:thermo} and \ref{subsec:charges} we work out the thermodynamic properties of the solution showing that the magnitude of the velocity acts like a chemical potential whose conjugate variable is the mass density. We further derive an Euler-type thermodynamic relation for Lifshitz perfect fluids using the conserved Noether charges and once more using the Killing charges associated with bulk Killing vectors. We end with a discussion of the first law of thermodynamics (summarized in \eqref{eq:intro},\eqref{eq:intro2}) for these Lifshitz black branes. 

Section \ref{subsec:WIs} can be read independently from sections \ref{sec:bulktheory} and \ref{sec:blackbranesmom}. It only requires appendix \ref{app:TNC}. It takes the point of view that these Lifshitz perfect fluids can be obtained by dimensional reduction of a relativistic perfect fluid (as discussed in appendix \ref{subsec:nullredfluid}) in the presence of a scalar source that is linearly dependent on the circle coordinate. It presents the Ward identities of a Lifshitz fluid and the expressions for the energy-momentum tensor and mass current at the perfect fluid level. The goal of this section is to derive the first law of our Lifshitz perfect fluids from the requirement that the Ward identities lead to the existence of a conserved entropy current. 
Finally, we present our conclusions and a number of open questions in Section \ref{sec:out}. 

\section{The bulk theory}\label{sec:bulktheory}

Lifshitz space-times as a solution of a theory with Einstein gravity coupled to matter fall into 2 classes. These are the Einstein--Proca-dilaton (EPD) theories of \cite{Gath:2012pg,Gouteraux:2012yr} and the Einstein--Maxwell dilaton (EMD) theories of \cite{Taylor:2008tg,Charmousis:2010zz}. We are interested in Lifshitz black brane solutions of the EPD models and to ultimately use them as a starting point to set up a fluid/gravity correspondence for Lifshitz space-times. Black brane solutions of theories with massive vectors were studied for models without a dilaton in \cite{Bertoldi:2009vn,Bertoldi:2009dt}. However in such theories the solutions are not analytically known. Nevertheless it is possible to work out the thermodynamics of these solutions by using first integrals of motion (with respect to the holographic radial coordinate) that allows one to relate near-horizon expansions to asymptotic expansions. We will follow a similar approach here.

Regarding Lifshitz black brane solutions of the EMD models, they are known analytically however they have different physical properties due to the presence of an extra bulk $U(1)$ symmetry and the fact that the dilaton is running logarithmically close to the boundary which introduces a new scaling exponent on top of the dynamical exponent $z$. The fluid/gravity correspondence for these black branes was studied in \cite{Kiritsis:2015doa}.

In this section we will introduce the EPD model and discuss its Lifshitz solutions. In the last section \ref{subsec:dimred} we will show that there is a specific EPD model that can be obtained from dimensional reduction of an action in one dimension higher that admits asymptotically AdS solutions. This so-called upliftable model will be used throughout the rest of this work.

\subsection{The EPD model}

The general class of 4-dimensional bulk EPD models is described by the following family of actions
\begin{equation}\label{eq:action}
 S=\int d^4x\sqrt{-g}\left(R-\frac{1}{4}Z(\Phi)F^2-\frac{1}{2}W(\Phi)B^2-\frac{x}{2}(\partial\Phi)^2-V(\Phi)\right)\,,
\end{equation}
where $F=dB$. The equations of motion are
\begin{align}
& G_{MN} = \frac{x}{2}\left(\partial_M\Phi\partial_N\Phi-
\frac{1}{2}(\partial\Phi)^2g_{MN}\right)-\frac{1}{2}V(\Phi)g_{MN}\nonumber\\
& +\frac{1}{2}Z(\Phi)\left(F_{MP}F_N{}^P-\frac{1}{4}F^2g_{MN}\right)+
\frac{1}{2}W(\Phi)(B_M B_N-\frac{1}{2}B^2g_{MN})\,,\label{eq:Einsteineq}\\
& \frac{x}{\sqrt{-g}}\partial_M\left(\sqrt{-g}\partial^M\Phi\right)=
\frac{1}{4}\frac{dZ}{d\Phi}F^2+\frac{1}{2}\frac{dW}{d\Phi}B^2+\frac{dV}{d\Phi}\,,\label{eq:scalareom}\\
& \frac{1}{\sqrt{-g}}\partial_M\left(\sqrt{-g}Z(\Phi)F^{MN}\right)=
W(\Phi)B^N\,.\label{eq:vectoreom}
\end{align}
The parameter $x$ can always be set equal to one but often it is more convenient to take some other value for it. It will prove convenient to make the following St\"uckelberg decomposition of the massive vector field
\begin{equation}\label{eq:axion}
B_M=A_M-\partial_M\Xi\,.
\end{equation}
The scalar $\Xi$ has dimensions of length and all other fields are dimensionless.

\subsection{Lifshitz solution}\label{subsec:Lifsol}

The equations of motion admit the following Lifshitz solutions (with $z>1$)
\begin{eqnarray}
 ds^2 & = & -\frac{1}{r^{2z}}dt^2+\frac{1}{r^2}\left(dr^2+dx^2+dy^2\right)\,,\label{eq:Lifshitz1}\\
 B & = & A_0\frac{1}{r^{z}}dt\,,\label{eq:Lifshitz2}\\
 \Phi & = & \Phi_\star\,,\label{eq:Lifshitz3}
\end{eqnarray}
provided that
\begin{eqnarray}
A_0^2 &=& \frac{2(z-1)}{z Z_0}\,,\label{eq:A0}\\
\frac{W_0}{Z_0} & = & 2z\,,\label{eq:m2}\\
 V_0 & = & -(z^2+z+4)\,,\label{eq:conditionV1}\\
V_1 & = & (za+2b)(z-1)\,.\label{eq:conditionV2}
\end{eqnarray}
where 
\begin{equation}\label{eq:def:a&b}
a=\frac{Z_1}{Z_0}\,,\qquad b=\frac{W_1}{W_0}\,.
\end{equation}
Above we have used the notation
\be
V_0\equiv V(\Phi_*)\sp V_1\equiv \frac{dV}{d\Phi} \Big|_{\Phi=\Phi_*}\sp V_2 \equiv \frac{d^2V}{d\Phi^2} \Big|_{\Phi=\Phi_*}
\ee
etc. where $ \Phi_\star$ is a constant. Equations \eqref{eq:m2} and \eqref{eq:conditionV1} determine the values of $\Phi_*$ and $z$ in terms of the functions appearing in the action. Equation \eqref{eq:A0} fixes $A_0$, and \eqref{eq:conditionV2} is a condition on the potential in order that Lifshitz is a solution of the family of actions \eqref{eq:action}. We note that there are also solutions of the EPD model with a logarithmically running scalar whose metric is a Lifshitz space-time \cite{Gouteraux:2012yr,Gath:2012pg}, but since these involve an additional scaling exponent these will not be considered here.

We can without loss of generality always perform a constant shift of $\Phi$ and redefine the functions $Z$, $W$ and $V$ such that for the new $\Phi$ the solution has $\Phi_\star=0$. We will from now on always assume this has been done.

In order to study the boundary fluid properties (or even only thermodynamic properties as we will do here) one needs to understand the near-boundary expansion and the identification in that expansion of all the sources and vevs. We will restrict our attention to a specific model for which this problem has been solved because it can be related to an AdS holographic renormalization problem in one dimension higher. The general features of Lifshitz black brane solutions of other EPD models will have to wait until we have understood fully the problem of performing holographic renormalization for asymptotically Lifshitz solutions of the general class of EPD models (see appendix \ref{subsec:comparison} for additional comments). The model for which we do have full control of the asymptotic expansion is called the upliftable model and this will be the subject of the next subsection.

\subsection{The upliftable model}\label{subsec:dimred}

When we make the choices
\begin{equation}\label{eq:upliftablemodel}
Z=e^{3\Phi}\,,\qquad W=4\,,\qquad V(\Phi)=2e^{-3\Phi}-12e^{-\Phi}\,,\qquad x=3\,.
\end{equation}
the action \eqref{eq:action} can be uplifted to
\begin{equation}\label{eq:uplifted}
S=\frac{1}{2\kappa_5^2}\int d^5x\sqrt{-\mathcal{G}}\left(R+12-\frac{1}{2}\partial_{\mathcal{M}}\psi\partial^{\mathcal{M}}\psi\right)\,,
\end{equation}
where $\kappa_5^2=8\pi G_5$ with $G_5$ the 5-dimensional Newton's constant and where $\mathcal{M}=(u,M)$. The relation between the 5- and 4-dimensional theories is via a so-called Scherk--Schwarz reduction whereby we demand that the scalar field $\psi$ when going around the compactification circle comes back to itself up to a shift. This is also known as a twisted reduction. This is possible because the higher-dimensional theory has a shift symmetry acting on the scalar. The Scherk--Schwarz reduction leading to \eqref{eq:action} with the choices \eqref{eq:upliftablemodel} is obtained by the following ansatz
\begin{eqnarray}
	ds_5^{2} &=& \mathcal{G}_{\mathcal{M}\mathcal{N}} dx^{\mathcal{M}}dx^{\mathcal{N}} = \frac{dr^{2}}{r^2} + \gamma_{AB}dx^{A}dx^{B} = e^{-\Phi}g_{MN}dx^{M}dx^{N} + e^{2\Phi}\left(du + A_{M}dx^{M}  \right)^2 \nonumber \\
	& =& e^{-\Phi}\left(e^{\Phi} \frac{dr^{2}}{r^2} + h_{\mu\nu}dx^{\mu}dx^{\nu}\right) + e^{2\Phi}\left(du + A_{\mu}dx^{\mu} \right)^2\,,\label{eq:red1} \\
	\psi &=& 2u+2\Xi\,, \label{eq:red2}
\end{eqnarray}
where the four dimensional fields $g_{MN}$, $A_M$, $\Xi$ and $\Phi$ are independent of the fifth coordinate $u$ which is periodically identified $u \sim u + 2\pi L$. In our normalization the 4-dimensional Newton's constant $G_4$ obeys $16\pi G_4=1$. This means that 5-dimensional Newton's constant $G_5$ obeys $\tfrac{2\pi L}{16\pi G_5}=\frac{1}{16\pi G_4}=1$.

What the Scherk--Schwarz reduction does is that it gauges the shift symmetry of $\psi$ using the Kaluza--Klein vector. In 4 dimensions this results in a covariant derivative acting on $\Xi$. This covariant derivative can be read as a massive vector field $B$ where $B$ is given by \eqref{eq:axion}. The consistency of the reduction will be proven in appendix \ref{subsec:consistency}. We now specialize to the case of the upliftable model \eqref{eq:upliftablemodel} because for this theory we have full control over the asymptotic solution space.

\section{Black branes with linear momentum}\label{sec:blackbranesmom}

The goal of this work is to construct the gravity dual of a Lifshitz perfect fluid. The Lifshitz algebra does not contain a boost generator. We will be interested in cases where the Lifshitz algebra arises from a larger algebra that contains boosts\footnote{These can only be Galilean or Carrollian boosts as these are compatible with a $z>1$ scaling. We cannot obtain a Lifshitz algebra by breaking Lorentz boosts because these require $z=1$.} by some explicit symmetry breaking. The bulk Lifshitz space-time has no boost Killing vector and the EPD model has no additional local symmetries that could combine with a space-time transformation to give an additional global symmetry like a Galilean boost\footnote{What we have in mind here is some bulk dual of the mechanism discussed in \cite{Hartong:2014pma,Hartong:2015wxa} whereby a boundary space-time transformations combined with a certain $U(1)$ transformation leads to an additional global symmetry. For example Galilean boost symmetries of the Schr\"odinger equation come about by a combination of a space-time Galilean coordinate transformation and a $U(1)$ phase transformation of the wave function. The latter can be traded for a $U(1)$ transformation of a background gauge field.}. Hence in order to study perfect fluids with a non-zero velocity we cannot simply boost a static Lifshitz black brane and promote the boost velocity to the fluid velocity. If we do that for a static Lifshitz black brane solution of the EPD model we simply describe a static black brane in a moving coordinate system and that is not equivalent to a moving black brane in a static coordinate system because of the absence of a boost symmetry. That means that we need to construct a new class of Lifshitz black branes that has a nonzero velocity or as we shall say nonzero linear momentum. We will construct these solutions near the Lifshitz boundary and near the horizon. We will then construct first integrals of motion to relate near-horizon quantities such as temperature and entropy to near-boundary quantities such as energy and mass density\footnote{If we assume that a Galilean boost symmetry has been broken, the velocity $V^i$ or rather by rotational invariance, its magnitude $V^2$, will be a chemical potential. On dimensional grounds it follows that the dual thermodynamic variable must be a mass density denoted by $\rho$. We will show that Lifshitz black branes indeed contain such a quantity.}. 

\subsection{The ansatz}\label{subsec:ansatz}

We assume that the black brane solution admits time and space translation Killing vectors. We can perform a rotation to make sure that the linear momentum is only along the $y$-direction. The full non-linear solution is thus of the form
\begin{eqnarray}
ds^2_4 & = &-F_1(r)\frac{dt^2}{r^{4}}+\frac{1}{F_2(r)}\frac{dr^2}{r^2}+F_3(r)\frac{dx^2}{r^2}+F_4(r)\left(\frac{dy}{r}+N(r)\frac{dt}{r^2}\right)^2\,,\label{eq:branewithP}\\
B & = & G_1(r)\frac{dt}{r^2} +G_2(r)\left(\frac{dy}{r}+N(r)\frac{dt}{r^2}\right)\,,\\
\Phi & = & \Phi(r)\,,\label{eq:branewithP3}
\end{eqnarray}
where we did not fix the $r$ reparametrization invariance. The powers in $r$ are chosen for convenience to match with the Lifshitz scaling of the boundary coordinates $t$, $x$ and $y$.

We can go to Eddington--Finkelstein coordinates\footnote{Null geodesics with generalized momenta $\frac{\partial L}{\partial\dot x}=0$ and $\frac{\partial L}{\partial\dot y}=0$ where $L=\frac{1}{2}g_{\mu\nu}\dot x^\mu \dot x^\nu$ correspond to $V=\text{cst}$ and $Y=\text{cst}$.} by defining $V$ and $Y$ coordinates as follows
\begin{eqnarray}
dt & = & dV+r\left(F_1F_2\right)^{-1/2}dr\,,\\
dy & = & dY-N\left(F_1F_2\right)^{-1/2}dr\,,
\end{eqnarray}
leading to 
\begin{eqnarray}
ds_4^2 & = & -F_1\frac{dV^2}{r^4}-2\left(\frac{F_1}{F_2}\right)^{1/2}\frac{dVdr}{r^{3}}+F_3\frac{dx^2}{r^2}+F_4\left(\frac{dY}{r}+N\frac{dV}{r^2}\right)^2\,,\label{eq:ansatzEF1}\\
B & = & G_1\frac{dV}{r^2}+\frac{G_1}{\left(F_1F_2\right)^{1/2}}\frac{dr}{r}+G_2\left(\frac{dY}{r}+N\frac{dV}{r^2}\right)\,.\label{eq:ansatzEF2}
\end{eqnarray}

Substituting the ansatz \eqref{eq:branewithP}--\eqref{eq:branewithP3} into the bulk equations of motion \eqref{eq:Einsteineq}--\eqref{eq:vectoreom} for \eqref{eq:upliftablemodel} and integrating the equations to an action leads to the following effective Lagrangian for the equations of motion 
\begin{eqnarray}
 L & = & r^{-5}\left(F_1F_2F_3F_4\right)^{1/2}\left[\frac{r^2}{2}\frac{F_1'}{F_1}\frac{F_3'}{F_3}+\frac{r^2}{2}\frac{F_1'}{F_1}\frac{F_4'}{F_4}-r\frac{F_3'}{F_3}-r\frac{F_4'}{F_4}+2r\frac{F_2'}{F_2}+\frac{r^2}{2}\frac{F_3'F_4'}{F_3F_4}-6\right.\nonumber\\
&&\left.+\frac{1}{2}\frac{F_4(rN'-N)^2}{F_1}+\frac{1}{2}\frac{Z\left(rG_1'-2G_1\right)^2}{F_1}+2\frac{G_1^2}{F_1F_2}-\frac{3}{2}r^2\Phi'^2-\frac{V}{F_2}-2\frac{G_2^2}{F_2F_4}\right.\label{eq:action3}\\
&&\left.+\frac{1}{2}\frac{ZG_2^2 \left(rN'-N\right)^2 }{F_1}-2\frac{ZG_1 G_2\left(rN'-N\right)}{F_1}+r\frac{Z G_2 G_1' \left(rN'-N\right)}{F_1}-\frac{1}{2}\frac{Z\left(rG_2'-G_2\right)^2}{F_4}\right]\,,\nonumber
\end{eqnarray}
where the independent functions are $F_1$ to $F_4$, $N$, $G_1$, $G_2$, $\Phi$ and their derivatives with respect to $r$. This effective Lagrangian can also be obtained by substituting the ansatz \eqref{eq:branewithP}--\eqref{eq:branewithP3} into the bulk action \eqref{eq:action} with \eqref{eq:upliftablemodel} and performing a few partial integrations. This ansatz is a generalization of a static black brane with zero momentum corresponding to setting $G_2=N=0$ and $F_3=F_4$.

The effective Lagrangian \eqref{eq:action3} has the following two scaling symmetries
\begin{equation}
F_1\rightarrow\lambda^2 F_1\,,\quad F_{3,4}\rightarrow\lambda^{-1}F_{3,4}\,,\quad N\rightarrow\lambda^{3/2} N\,,\quad G_1\rightarrow\lambda G_1\,,\quad G_2\rightarrow\lambda^{-1/2}G_2\,,
\end{equation}
and 
\begin{equation}
F_3\rightarrow\mu^2 F_3\,,\quad F_4\rightarrow\mu^{-2}F_4\,,\quad N\rightarrow\mu N\,,\quad G_2\rightarrow\mu^{-1}G_2\,.
\end{equation}
Both of these transformations are symmetries of the ansatz \eqref{eq:branewithP}--\eqref{eq:branewithP3} provided we transform the coordinates appropriately. For the $\lambda$ transformation that means that we must rescale the coordinates as
\begin{equation}\label{eq:global1}
t\rightarrow\lambda^{-1}t\,,\qquad x\rightarrow\lambda^{1/2}x\,,\qquad y\rightarrow\lambda^{1/2}y\,,
\end{equation}
while for the $\mu$ transformation it means that we must rescale the spatial coordinates as
\begin{equation}\label{eq:global2}
x\rightarrow\mu^{-1}x\,,\qquad y\rightarrow\mu y\,.
\end{equation}
Using Noether's theorem the associated charges are $Q_\lambda$ and $Q_\mu$, respectively, that are given by
\begin{eqnarray}
Q_\lambda & = & -2\frac{\partial L}{\partial F_1'}F_1+\frac{\partial L}{\partial F_3'}F_3+\frac{\partial L}{\partial F_4'}F_4-\frac{3}{2}\frac{\partial L}{\partial {N}'}N-\frac{\partial L}{\partial G_1'}G_1+\frac{1}{2}\frac{\partial L}{\partial G_2'}G_2\,,\\
Q_\mu & = & -2\frac{\partial L}{\partial F_3'}F_3+2\frac{\partial L}{\partial F_4'}F_4-\frac{\partial L}{\partial {N}'}N+\frac{\partial L}{\partial G_2'}G_2\,.
\end{eqnarray}
Using that $L$ is given by \eqref{eq:action3} these charges can be shown to be equal to
\begin{eqnarray}
\hspace{-.9cm}Q_\lambda & = & -\frac{3}{2}r^{-1}Q_N N+r^{-4}\left(F_1F_2F_3F_4\right)^{1/2}\left[-\frac{ZG_1}{F_1}\left(rG_1'-2G_1\right)-2+r\frac{F_1'}{F_1}-\frac{r}{2}\frac{F'_3}{F_3}-\frac{r}{2}\frac{F'_4}{F_4}\right.\nonumber\\
\hspace{-.9cm}&&\left.-\frac{ZG_1G_2\left(rN'-N\right)}{F_1}-\frac{1}{2}\frac{ZG_2\left(rG_2'-G_2\right)}{F_4}\right]\,,\label{eq:Qlambda}
\end{eqnarray}
and
\begin{eqnarray}
Q_\mu & = & -r^{-1}Q_N N+r^{-4}\left(F_1F_2F_3F_4\right)^{1/2}\left[r\frac{F_3'}{F_3}-r\frac{F_4'}{F_4}-\frac{ZG_2\left(rG_2'-G_2\right)}{F_4}\right]\,,\label{eq:Qnu}
\end{eqnarray}
where we defined the charge $Q_N$
\begin{eqnarray}
Q_N & = & \frac{\partial L}{\partial {N}'}=r^{-3}\left(F_1F_2F_3F_4\right)^{1/2}\left[\frac{F_4\left(rN'-N\right)}{F_1}+\frac{ZG_2^2\left(rN'-N\right)}{F_1}-2\frac{ZG_1G_2}{F_1}\right.\nonumber\\
&&\left.+\frac{ZG_2\left(rG_1'-G_1\right)}{F_1}\right]\,,
\end{eqnarray}
which results from the fact that $L$ does not depend on $N$. The Noether charges $Q_\lambda$ and $Q_\mu$ are first integrals of motion and thus independent of the radial coordinate $r$. This will play an important role later when we derive the thermodynamic properties. We will see that $Q_\lambda$ relates to the energy and $Q_\mu$ to the linear momentum of the black brane.

The ansatz \eqref{eq:branewithP}--\eqref{eq:branewithP3} has a third global scale symmetry namely
\begin{equation}\label{eq:global3}
t\rightarrow\nu^{-1}t\,,\qquad F_1\rightarrow\nu^2 F_1\,,\qquad N\rightarrow\nu N\,,\qquad G_1\rightarrow\nu G_1\,.
\end{equation}
However this transformation does not leave the effective Lagrangian \eqref{eq:action3} invariant because it is not a symmetry of the prefactor. On top of the 3 global symmetries whose parameters are $\lambda$, $\mu$ and $\nu$ the ansatz also has one local symmetry which corresponds to $r$-reparametrization invariance. This symmetry acts as
\begin{equation}
\delta F_2=\xi^rF_2'+2F_2\left(r^{-1}\xi^r-\partial_r\xi^r\right)\,,\qquad \delta A_I=\xi^r A_I'\,,
\end{equation}
where $A_I$ is any of the functions appearing in the ansatz that is not $F_2$ and $\xi^r$ is the local parameter generating the $r$-reparametrization. This local symmetry can be fixed by choosing a gauge. This local symmetry implies that using the $F_2$ equation of motion (which is first order and needs to be differentiated with respect to $r$) and any 6 of the other $A_I$ equations of motion the remaining 7th $A_I$ equation of motion can be derived. 

\subsection{The asymptotic solution}\label{subsec:asympsol}

The 4-dimensional near-boundary expansion follows by dimensional reduction using the reduction ansatz \eqref{eq:red1} and \eqref{eq:red2} as well as the 5-dimensional Fefferman--Graham (FG) expansion \eqref{eq: sol metric} and \eqref{eq: sol phi}, the details of which are given in appendix \ref{sec:FGexpansions}.

The ansatz for the black branes with linear momentum are such that all 4-dimensional fields only depend on the radial coordinate $r$. From the 5-dimensional FG expansion point of view that implies that all sources and vevs must be constants. The only exception to this is of course the fact that $\psi$ is allowed to be linear in the reduction circle coordinate $u$ because we are performing a Scherk--Schwarz reduction. That means that our ansatz forces us to consider a FG expansion in 5D with the following sources and vevs
\begin{eqnarray}
\gamma_{(0)AB} & = & \text{cst}\,,\qquad\text{with $\gamma_{(0)uu}=0$}\,,\\
t_{AB} & = & \text{cst}\,,\\
\psi_{(0)} & = & 2u\,,\qquad\langle O_\psi\rangle = 0\,,\qquad\text{so that $\Xi=0$}\,,
\end{eqnarray}
where $t_{AB}$ obeys the Ward identities \eqref{eq:tracet} and \eqref{eq:divt}. Setting $\langle O_\psi\rangle=0$ is a consequence of the Ward identity $\nabla_{(0)A} t^A{}_B=-\langle O_\psi\rangle\partial_B\psi$ for $B=u$ and constant $t^A{}_B$. Since the field $\Xi$ always appears differentiated it makes no difference if we set it equal to zero or equal to some constant. The choice $\gamma_{(0)uu}=0$ is rather important and is necessary in order that the lower-dimensional theory has a $z=2$ scaling exponent. This is explained in detail in \cite{Chemissany:2012du,Christensen:2013rfa}. It is shown in section 2 of \cite{Christensen:2013rfa} that the reduction in the bulk is everywhere along a spacelike circle (due to $\psi_{(0)}=2u$) but that this circle is null on the boundary\footnote{Here we use a model that is simpler than the one used in \cite{Christensen:2013rfa} but regarding this point the properties are identical.}.

It is well known that reductions along null Killing directions turn a Riemannian geometry into a torsional Newton--Cartan (TNC) geometry \cite{Eisenhart,Kuenzle:1972zw,Julia:1994bs,Christensen:2013rfa}. For details see appendices \ref{subsec:nullred} and \ref{subsec:sourcesvevs}. In particular see the reduction ansatz for the AdS$_5$ boundary metric \eqref{eq:higherDmetric}. In the language of TNC geometry the $uu$ component of the inverse metric is called $\tilde\Phi$ which is defined in \eqref{eq:tildePhi} with $m_\mu$ the Kaluza--Klein vector associated with the null reduction as given in \eqref{eq:higherDmetric}. In appendix \ref{subsec:sourcesvevs} it is shown that $m_\mu$ combines with $\chi$ the source of the bulk scalar $\Xi$ into $M_\mu=m_\mu-\partial_\mu\chi$. Since here we have $\chi=0$ we can take $M_\mu=m_\mu$. From the inverse metric we know that%
\footnote{We warn the reader that the boundary background field $ \tilde\Phi$ should not be confused with the bulk scalar field $\Phi$.}
\begin{equation}
\gamma_{(0)}^{uu}=2\tilde\Phi\,.
\end{equation}

We will be interested in flat boundaries of the 4-dimensional $z=2$ Lifshitz space-time. A flat space-time in TNC language means that there exists a coordinate system in which we have \cite{Hartong:2015wxa}
\begin{equation}\label{eq:flatTNC}
\tau_\mu=\delta_\mu^t\,,\qquad M_\mu=0\,,\qquad h_{tt}=h_{ti}=0\,,\qquad h_{ij}=\delta_{ij}\,.
\end{equation}
This means in particular that $\tilde\Phi=0$. Turning on $\tilde\Phi$ corresponds to turning on a Newtonian potential for the boundary theory \cite{Bergshoeff:2014uea,Hartong:2015wxa}. We will thus not consider this possibility.

The expansion of the 4-dimensional fields follows from \eqref{eq:red1} and \eqref{eq:red1} which imply that\footnote{We warn the reader that we use $h_{\mu\nu}$ both to denote the $\mu\nu$ component of the bulk metric as well as the spatial metric-like quantity \eqref{eq:h} on the boundary. We hope that this will not cause any confusion.}
\begin{eqnarray}
e^{2\Phi} & = & \gamma_{uu}\,,\label{eq:reduc1}\\
A_\mu & = & \frac{\gamma_{u\mu}}{\gamma_{uu}}\,,\label{eq:reduc2}\\
h_{\mu\nu} & = & \left(\gamma_{uu}\right)^{1/2}\left(\gamma_{\mu\nu}-\frac{\gamma_{u\mu}\gamma_{u\nu}}{\gamma_{uu}}\right)\,,\label{eq:reduc3}
\end{eqnarray}
where $\gamma_{AB}$ is FG expanded using the results of appendix \ref{sec:FGexpansions}. In order to carry out this reduction we need to know how to reduce the AdS boundary energy-momentum tensor into the language of the energy-momentum tensor of the TNC boundary of the lower-dimensional Lifshitz space-time. The relation between a relativistic energy-momentum tensor  $t_{MN}$ and the TNC energy-momentum tensor related via null reduction is explained in appendix \ref{subsec:nullredEMT} where we derive the following relations
\begin{eqnarray}
t_{uu} & = & \rho\,,\label{eq:5DEMT1}\\
t_{u\mu} & = & \tau_\rho T^\rho{}_\mu\,,\\
t_{\mu\nu} & = & \hat h_{\mu\rho}\hat h_{\nu\kappa}h^{\kappa\sigma}T^\rho{}_\sigma-\left(\tau_\nu\hat h_{\mu\rho}+\tau_\mu\hat h_{\nu\rho}\right)\hat v^\sigma T^\rho{}_\sigma+\left(\hat v^\rho\hat v^\sigma t_{\rho\sigma}\right)\tau_\mu\tau_\nu\,,\label{eq:5DEMT3}
\end{eqnarray}
where 
\begin{equation}
\hat v^\rho\hat v^\sigma t_{\rho\sigma}=t^{uu}-4\tilde\Phi^2\rho+4\tilde\Phi\hat v^\sigma\tau_\rho T^\rho{}_\sigma\,.
\end{equation}
Recall that here $\tilde\Phi=0$. The TNC energy-momentum tensor is denoted by $T^\mu{}_\nu$ and the TNC mass density is denoted by $\rho$. We note that $t^{uu}$ has no lower dimensional interpretation in terms of energy-momentum or mass density. As shown in appendix \ref{subsec:nullred} it does not appear in any of the Ward identities involving $T^\mu{}_\nu$ and $\rho$. Hence we will set $t^{uu}$ equal to zero. It would appear in 4 dimensions for the first time at order $r^2$ in that part of the expansion of $h_{\mu\nu}$ that is proportional to $\tau_\mu\tau_\nu$. We refer to \cite{Christensen:2013rfa} for more discussion on the role of $t^{uu}$.

In order to find out where the momentum flux, the spatial projection of $\tau_\rho T^\rho{}_\mu$, which is one of the quantities of interest, the spatial stress tensor etc. appear upon reduction we need to know what happens with $t_{\mu\nu}$ upon reduction. Clearly in 5 bulk dimensions $t_{\mu\nu}$ appears in $\gamma_{\mu\nu}$ at order $r^2$. Therefore in order to see it in four bulk dimensions we need to expand $A_\mu$ and $h_{\mu\nu}$ up to order $r^2$. This follows from \eqref{eq:reduc2} and \eqref{eq:reduc3} and implies that we need to expand $\gamma_{uu}$ to order $r^6$, $\gamma_{u\mu}$ to order $r^4$ and $\gamma_{\mu\nu}$ to order $r^2$. 

We will now proceed to construct the 5-dimensional solution up to the required order. Since the sources and vevs are constants (with $\psi_{(0)}$ linear in $u$) we have that $\square\psi=0$ implies
\begin{equation}
\sqrt{-\mathcal{G}}r^2\partial_r\psi=C\,,
\end{equation}
where $C$ is an integration constant. Using
\begin{equation}
\sqrt{-\mathcal{G}}=1+O(r^6)\,,
\end{equation}
it follows that
\begin{equation}
r^{-3}\partial_r\psi=C\left(1+O(r^6)\right)\,,
\end{equation}
so that using $\psi_{(4)}=0$ implies that $C=0$. Hence $\partial_r\psi=0$, or in other words $\psi=2u$ to all orders. With this result the Einstein equation simplifies to 
\begin{equation}
G_{\mathcal{M}\mathcal{N}} = 6\mathcal{G}_{\mathcal{M}\mathcal{N}}+2\delta_{\mathcal{M}}^u\delta_{\mathcal{N}}^u-\mathcal{G}_{\mathcal{M}\mathcal{N}}\gamma^{uu}\,,
\end{equation}
which is equivalent to
\begin{equation}\label{eq:Einsteqsimplified}
R_{\mathcal{M}\mathcal{N}} = -4\mathcal{G}_{\mathcal{M}\mathcal{N}}+2\delta_{\mathcal{M}}^u\delta_{\mathcal{N}}^u\,.
\end{equation}

To find the solution up to order $r^6$ we make the following ansatz
\begin{equation}
\gamma_{AB}=r^{-2}\left(\gamma_{(0)AB}+r^2\delta_A^u\delta_B^u-\frac{1}{4}r^4t_{AB}+r^6\gamma_{(6)AB}+r^8\gamma_{(8)AB}+O(r^{10})\right)\,.
\end{equation}
The log terms at order $r^2\log r$ are zero and so it is expected that they are zero to all orders. This is a correct assumption as long as we do not need to put constraints on the sources and vevs coming from the nature of the expansion. The inverse metric reads
\begin{equation}
\gamma^{AB}=r^2\left(\gamma_{(0)}^{AB}-r^2\gamma_{(0)}^{Au}\gamma_{(0)}^{Bu}+\frac{1}{4}r^4 t^{AB}+r^6\sigma_{(6)}^{AB}+r^8\sigma_{(8)}^{AB}+O(r^{10})\right)\,,
\end{equation}
where
\begin{eqnarray}
\sigma_{(6)}^{AB} & = & -\gamma_{(6)}^{AB}-\frac{1}{4}\gamma_{(0)}^{Au}t^{uB}-\frac{1}{4}\gamma_{(0)}^{Bu}t^{uA}\,,\\
\sigma_{(8)}^{AB} & = & -\gamma_{(8)}^{AB}+\gamma_{(0)}^{Au}\gamma_{(6)}^{uB}+\gamma_{(0)}^{Bu}\gamma_{(6)}^{uA}+\frac{1}{16}t^{AC}t_C{}^B+\frac{1}{4}\gamma_{(0)}^{Au}\gamma_{(0)}^{Bu}t^{uu}\,.
\end{eqnarray}
The Christoffel symbols are
\begin{eqnarray}
&&\Gamma^r_{rr} = -\frac{1}{r}\,,\qquad\Gamma^r_{rA} = 0\,,\qquad\Gamma^r_{AB} = -\frac{1}{2}r^2\partial_r\gamma_{AB}\,,\nonumber\\
&&\Gamma^A_{rr} = 0\,,\qquad\Gamma^A_{rB} = \frac{1}{2}\gamma^{AC}\partial_r\gamma_{BC}\,,\qquad\Gamma^A_{BC} = 0\,.
\end{eqnarray}
From this we conclude that 
\begin{equation}
R_{rr}=-4r^{-2}+r^4\left(-12\gamma_{(6)A}^A-2t^{uu}\right)+r^6\left(-24\gamma_{(8)A}^A+18\gamma_{(6)}^{uu}+\frac{1}{2}t^{AB}t_{AB}\right)+O(r^8)\,.
\end{equation}
The $rr$ component of \eqref{eq:Einsteqsimplified} tells us that $R_{rr}=-4r^{-2}$ so that
\begin{equation}
\gamma_{(6)A}^A = -\frac{1}{6}t^{uu}\,,\qquad\gamma_{(8)A}^A = \frac{3}{4}\gamma_{(6)}^{uu}+\frac{1}{48}t^{AB}t_{AB}\,.
\end{equation}
The $rA$ component of \eqref{eq:Einsteqsimplified} brings nothing as both sides are identically zero. Using that
\begin{eqnarray}
R_{AB} & = & -4r^{-2}\gamma_{(0)AB}-2\delta_A^u\delta_B^u+r^2t_{AB}+r^4\left(\frac{1}{4}t^{uu}\gamma_{(0)AB}-10\gamma_{(6)AB}-\frac{1}{2}\delta_A^u t^u{}_B-\frac{1}{2}\delta_B^u t^u{}_A\right)\nonumber\\
&&+r^6\left(-20\gamma_{(8)AB}-\gamma_{(6)}^{uu}\gamma_{(0)AB}-\frac{1}{24}t^{CD}t_{CD}\gamma_{(0)AB}+6\delta_A^u\gamma_{(6)B}^u+6\delta_B^u\gamma_{(6)A}^u\right.\nonumber\\
&&\left.+\frac{1}{2}t_A{}^Ct_{CB}+\frac{1}{2}\delta_A^u\delta_B^u t^{uu}\right)+O(r^8)\,,
\end{eqnarray}
as well as the equation of motion $R_{AB}=-4\gamma_{AB}+2\delta_A^u\delta_B^u$, we find that
\begin{eqnarray}
\gamma_{(6)AB} & = & -\frac{1}{6}\delta^u_A t^u{}_B-\frac{1}{6}\delta^u_B t^u{}_A+\frac{1}{24}t^{uu}\gamma_{(0)AB}\,,\\
\gamma_{(8)AB} & = & -\frac{1}{16}t^{uu}\delta^u_A\delta^u_B-\frac{1}{384}t^{CD}t_{CD}\gamma_{(0)AB}+\frac{1}{32}t_A{}^Ct_{BC}\,.
\end{eqnarray}

From the reduction \eqref{eq:reduc1}--\eqref{eq:reduc3} it follows that
\begin{eqnarray}
\Phi & = & -\frac{1}{8}r^2\rho+r^4\left(\frac{1}{6}\hat v^\sigma\tau_\rho T^\rho{}_\sigma-\frac{1}{64}\rho^2\right)+O(r^6)\,,\\
A_\mu & = & r^{-2}\tau_\mu+\frac{1}{4}\rho\tau_\mu+r^2\left(\frac{1}{12}\tau_\rho T^\rho{}_\mu+\frac{1}{16}\rho^2\tau_\mu-\frac{1}{3}\bar h_{\mu\rho}T^\rho\right)+O(r^4)\,,\\
h_{\mu\nu} & = & -r^{-4}\tau_\mu\tau_\nu+r^{-2}\left(\bar h_{\mu\nu}-\frac{1}{8}\rho\tau_\mu\tau_\nu\right)-\frac{1}{8}\rho\bar h_{\mu\nu}+\frac{1}{4}\left(\tau_\mu\tau_\rho T^\rho{}_\nu+\tau_\nu\tau_\rho T^\rho{}_\mu\right)\nonumber\\
&&-\left(\frac{3}{128}\rho^2-\frac{1}{6}\hat v^\sigma\tau_\rho T^\rho{}_\sigma\right)\tau_\mu\tau_\nu+r^2\left(-\frac{1}{4}\hat h_{\mu\rho}\hat h_{\nu\kappa}h^{\kappa\sigma}T^\rho{}_\sigma+\frac{1}{12}\left(\tau_\nu\hat h_{\mu\rho}+\tau_\mu\hat h_{\nu\rho}\right)\hat v^\sigma T^\rho{}_\sigma\right.\nonumber\\
&&\left.+\left(\frac{1}{6}\hat v^\sigma\tau_\rho T^\rho{}_\sigma-\frac{1}{128}\rho^2\right)\bar h_{\mu\nu}-\frac{1}{16}\rho\left(\tau_\mu\tau_\rho T^\rho{}_\nu+\tau_\nu\tau_\rho T^\rho{}_\mu\right)\right.\nonumber\\
&&\left.+\left(\frac{3}{64}\rho\hat v^\sigma\tau_\rho T^\rho{}_\sigma+\frac{1}{64}T^\sigma\tau_\rho T^\rho{}_\sigma-\frac{5}{1024}\rho^3-\frac{1}{32}t^{uu}\right)\tau_\mu\tau_\nu\right)+O(r^4)\,.
\end{eqnarray}
For the interested reader we have included the term $t^{uu}$. But, as remarked earlier, we will set this independent quantity equal to zero. If we choose the boundary sources to correspond to a flat TNC boundary as in \eqref{eq:flatTNC} then the expansions become
\begin{eqnarray}
\Phi & = & -\frac{1}{8}r^2\rho-r^4\left(\frac{1}{6}T^t{}_t+\frac{1}{64}\rho^2\right)+O(r^6)\,,\label{eq:asymp1}\\
A_t & = & r^{-2}+\frac{1}{4}\rho+r^2\left(\frac{1}{12}T^t{}_t+\frac{1}{16}\rho^2\right)+O(r^4)\,,\\
A_i & = & -\frac{1}{4}r^2 T^t{}_i+O(r^4)\,,\\
h_{tt} & = & -r^{-4}-\frac{1}{8}r^{-2}\rho+\frac{1}{3}T^t{}_t-\frac{3}{128}\rho^2+O(r^2)\,,\\
h_{ti} & = & \frac{1}{4}T^t{}_i+r^2\left(-\frac{1}{12}\delta_{ij}T^j{}_t+\frac{1}{32}\rho T^t{}_i\right)+O(r^4)\,,\\
h_{ij} & = & r^{-2}\delta_{ij}-\frac{1}{8}\rho\delta_{ij}+r^2\left(\left(\frac{1}{12}T^t{}_t-\frac{1}{128}\rho^2\right)\delta_{ij}-\frac{1}{4}\delta_{ik}T^k{}_j+\frac{1}{8}T^k{}_k\delta_{ij}\right)+O(r^4)\,,\label{eq:asymp6}
\end{eqnarray}
where in the last expression we used the $z$-deformed trace Ward identity (equation \eqref{eq:nullredtraceWI} with zero on the right hand side)
\begin{equation}\label{eq:z=2trace}
2T^t{}_t+T^k{}_k=0\,.
\end{equation}

In this work we are interested in gravitational duals of boundary perfect fluids so without loss of generality we can assume that $T^\mu{}_\nu$ takes the form of a perfect fluid. This form is derived in appendix \ref{subsec:nullredfluid} by the null reduction of a relativistic perfect fluid. On flat TNC space-time it reads
\begin{eqnarray}
&&T^t{}_t = -\left(\mathcal{E}+\frac{1}{2}\rho V^2\right)\,,\qquad T^i{}_t = -\left(\mathcal{E}+P+\frac{1}{2}\rho V^2\right)V^i\,,\label{eq:bdryPF1}\\
&&T^t{}_i = \rho V_i\,,\qquad T^j{}_i = \left(P\delta^j_i+\rho V^j V_i\right)\,,\label{eq:bdryPF2}
\end{eqnarray}
where $\mathcal{E}$ is the energy density, $P$ the pressure, $\rho$ the mass density and $V^i$ the velocity of the fluid. The $z$-deformed trace Ward identity tells us that the equation of state is $P=\mathcal{E}$.

It is interesting and insightful to take a closer at look at this $V$-dependent solution from the 5-dimensional point of view. Using the relations between the lower and higher-dimensional energy-momentum tensors \eqref{eq:5DEMT1}--\eqref{eq:5DEMT3} we see that the 5-dimensional energy-momentum is given by
\begin{eqnarray}
&& t_{uu}=\rho\,,\quad t_{ut}=-\mathcal{E}-\frac{1}{2}\rho V^2\,,\quad t_{ui}=\rho V_i\,,\nonumber\\
&& t_{ti}=-\left(\mathcal{E}+P+\frac{1}{2}\rho V^2\right)V_i\,,\quad t_{ij}=P\delta_{ij}+\rho V_i V_j\,,
\end{eqnarray}
with $t_{tt}$ being undetermined. A convenient way of writing this is in terms of $t_{AB}dx^A dx^B$, which can be seen to be equal to
\begin{eqnarray}
t_{AB}dx^A dx^B & = & \rho\left(du+V_i dx^i-\frac{1}{2}V^2dt\right)^2-2\mathcal{E}dt\left(du+V_i dx^i-\frac{1}{2}V^2dt\right)\\
&&+P\delta_{ij}\left(dx^i-V^idt\right)\left(dx^j-V^jdt\right)+\left(t_{tt}-\left(\mathcal{E}+P+\frac{1}{2}\rho V^2\right)V^2\right)dt^2\nonumber\,.
\end{eqnarray}
The rest of the solution is fully determined by the following boundary data
\begin{eqnarray}
\gamma_{(0)AB}dx^A dx^B & = & 2dtdu+\delta_{ij}dx^i dx^j\,,\\
\psi_{(0)} & = & 2u\,,\qquad\langle O_\psi\rangle = 0\,.
\end{eqnarray}
If we now perform the following coordinate transformation, which from a lower dimensional point of view is a Galilean boost and a $U(1)$ gauge transformation (acting on the Kaluza--Klein vector $m_\mu$), 
\begin{equation}
u = u'-\frac{1}{2}V^2t'-V_i x'^i\,,\qquad t =t'\,,\qquad x^i=x'^i+V^i t'\,,
\end{equation}
we obtain
\begin{eqnarray}
t_{AB}dx^A dx^B & = & \rho du'^2-2\mathcal{E}dt' du'+P\delta_{ij}dx'^idx'^j\nonumber\\
&&+\left(t_{tt}-\left(\mathcal{E}+P+\frac{1}{2}\rho V^2\right)V^2\right)dt'^2\,,\\
\gamma_{(0)AB}dx^A dx^B & = & 2dt'du'+\delta_{ij}dx'^i dx'^j\,,\\
\psi_{(0)} & = & 2u'-V^2t'-2V_ix'^i\,,\\
\langle O_\psi\rangle & = & 0\,.
\end{eqnarray}
We thus see that the boundary metric $\gamma_{(0)AB}$ remained invariant and that all the $V$-dependence now resides in the expression for $\psi_{(0)}$. The $t't'$ component of $t_{AB}$ is not important for the lower dimensional boundary energy-momentum tensor and its Ward identities. It is thus clear that due to the presence of $\psi$, and the Scherk--Schwarz reduction ansatz $\psi=2u+\Xi$, solutions with different $V^i$ are not diffeomorphic. We will later see this reflected in the fact that $V^2$ plays the role of a chemical potential. The ansatz in section \ref{subsec:ansatz} used rotations to orient the flow in the $y$-directions. We will see further below that indeed $V^x=0$.

\subsection{The near-horizon solution}

The near-horizon expansion is entirely straightforward. Referring to the ansatz \eqref{eq:ansatzEF1} and \eqref{eq:ansatzEF2} in EF coordinates we can make the following observations about the behavior of the solution near the horizon.

The horizon is located at the locus where the $r=\text{cst}$ hypersurface becomes null, i.e. at $g^{rr}=0$. That means that $F_2$ will have a first order zero at $r=r_h$. Regularity of the metric in EF coordinates, in particular of the component $g_{Vr}$ then tells us that $F_1$ must also have a first order zero at $r=r_h$. Note that for $N\neq 0$ this is not the locus where $\partial_t$ becomes null. In other words the stationary limit surface $g_{tt}=0$ comes before the horizon (viewed from outside). Regularity of the massive vector at the horizon forces $G_1$ to have a first order zero at $r=r_h$. The functions $F_4$ and $N$ are both regular without any zeros at the horizon, i.e. $F_4(r_h)\neq 0$ and $N(r_h)\neq 0$. The latter quantity can be zero but as we will see in the next subsection that corresponds to a brane without any momentum so we take it to be nonzero. The remaining functions $G_2$ and $\Phi$ are regular at the horizon, but they do not have to be non-vanishing.

A convenient gauge choice to fix the $r$ reparametrization invariance of the ansatz to study the near-horizon horizon solution is to take $F_3=1$. In this gauge we will refer to the radial coordinate as $R$ to distinguish it from the radial coordinate $r$ used in the previous subsection\footnote{We permit ourselves to also use $r$ for the family of gauges parametrized by the ansatz \eqref{eq:ansatzEF1} and \eqref{eq:ansatzEF2}. We hope that this will not cause any confusion.}. The horizon is now located at $R=R_h$.

The ansatz also has three global scale symmetries \eqref{eq:global1}, \eqref{eq:global2} and \eqref{eq:global3} that leave the ansatz invariant. These can be viewed as rescalings of $x$, $y$ and $t$. We have used these symmetries to set the asymptotic values of $F_1$, $F_3$ and $F_4$ equal to one. This fixes the asymptotic values of $\Phi$ and thus of $F_2$ (via the asymptotic gauge choice $F_2=e^{-\Phi}$) as well as of $G_1$ via the equations of motion. That means that we cannot use these rescaling symmetries a second time to fix parameters in the near-horizon solution. We thus take for the near-horizon solution the following expansion
\begin{eqnarray}
F_1 & = & f_1\frac{R-R_h}{R_h}+\ldots\,,\label{eq:nearhor1}\\
F_2 & = & h_1\frac{R-R_h}{R_h}+\ldots\,,\\
F_3 & = & 1\,,\\
F_4 & = & p_0+p_1\frac{R-R_h}{R_h}+\ldots\,,\\
N & = & n_0+n_1\frac{R-R_h}{R_h}+\ldots\,,\\
G_1 & = & g_1\frac{R-R_h}{R_h}+\ldots\,,\\
G_2 & = & m_0+m_1\frac{R-R_h}{R_h}+\ldots\,,\\
\Phi & = & l_0+l_1\frac{R-R_h}{R_h}+\ldots\,.\label{eq:nearhor8}
\end{eqnarray}

Most but not all of the coefficients appearing in the near-horizon expansion will be determined by solving the equations of motion of the effective action $L$ in an expansion around $R=R_h$. We studied the solution up to second order in $R-R_h$ and it leaves 8 parameters unfixed. These are $f_1$, $p_0$, $g_1$, $m_0$, $n_0$, $n_1$, $l_0$ and $r_h$. The parameter $h_1$ is fixed by the equations of motion to be\footnote{To find this result one solves the leading term of the $F_1$ equation of motion for $p_1$ and the leading term of the $F_3$ equation of motion for $f_2$. The expression then follows from the leading term in the $F_4$ equation of motion. A similar expression has been observed in \cite{Bertoldi:2009vn}.} 
\begin{equation}
h_1=\frac{2f_1\left(2e^{-3l_0}-12 e^{-l_0}\right)}{4f_1-e^{3l_0}\left(g_1+m_0n_1\right)^2}\,,
\end{equation}
where the numerator is $2f_1$ times the potential \eqref{eq:upliftablemodel} evaluated at $R=R_h$. We expect that most of these parameters will be determined by matching the solution onto the asymptotic region.

There are not many examples known of analytic black brane solutions of the EPD model. However in the context of Schr\"odinger space-times we can obtain analytic solutions by applying a sequence of duality transformations known as TsT transformations \cite{Maldacena:2008wh} to obtain black brane solutions from known AdS black branes \cite{Adams:2008wt,Herzog:2008wg,Hartong:2010ec}. The resulting Schr\"odinger black branes have a nonzero charge associated with particle number. Since in Schr\"odinger holography particle number is realized geometrically this means that these correspond to black branes with a linear momentum along a direction that asymptotically becomes null. If we study these black branes near the horizon in the same coordinates in which the AdS black brane has a flat boundary Minkowski metric written in Cartesian coordinates then we see the exact same near-horizon boundary conditions as we imposed for our Lifshitz black brane\footnote{More explicitly if we use equation (62) of \cite{Hartong:2010ec} setting $\xi=V$ the TsT transformation (113)--(115) provides us with a $z=2$ Schr\"odinger black brane solution of some EPD model. If we then perform the coordinate transformation $t=T-X$ and $2\xi=2V=T+X$ we find that the near-horizon geometry has exactly the same properties as the Lifshitz black brane solution studied here.}.

\subsection{Comments on the interpolating solution}\label{subsec:interpolating}

We have used different radial gauges in the near-horizon region ($F_3=1$) and in the asymptotic region ($F_2=e^{-\Phi}$). The two coordinates are related via the coordinate transformation 
\begin{equation}\label{eq:changegauge}
h_{xx}=R^{-2}\,,
\end{equation}
where $h_{xx}$ is given in \eqref{eq:asymp6}. In order to write both the near-horizon and the near-boundary expansion in the same gauge it is convenient to rewrite the expansions \eqref{eq:asymp1}--\eqref{eq:asymp6} in terms of the radial coordinate $R$. This can be done as follows. The expansions \eqref{eq:asymp1}--\eqref{eq:asymp6} in terms of the ansatz functions correspond to
\begin{eqnarray}
F_1 & = & 1+\frac{1}{8}r^2\rho +r^4\left(-\frac{1}{3}T^t{}_t+\frac{3}{128}\rho^2\right)+O(r^6)\,,\\
F_2 & = & e^{-\Phi}=1+\frac{1}{8}r^2\rho +r^4\left(\frac{1}{6}T^t{}_t+\frac{3}{128}\rho^2\right)+O(r^6)\,,\\
F_3 & = & 1-\frac{1}{8}r^2\rho +r^4\left(\frac{1}{12}T^t{}_t-\frac{1}{128}\rho^2+\frac{1}{8}\rho V^2\right)+O(r^6)\,,\\
F_4 & = & 1-\frac{1}{8}r^2\rho +r^4\left(\frac{1}{12}T^t{}_t-\frac{1}{128}\rho^2-\frac{1}{8}\rho V^2\right)+O(r^6)\,,\\
N & = & \frac{1}{4}r^3\rho V+O(r^5)\,,\\
G_1 & = & 1+\frac{1}{4}r^2\rho +r^4\left(\frac{1}{12}T^t{}_t+\frac{1}{16}\rho^2\right)+O(r^6)\,,\\
G_2 & = & -\frac{1}{4}r^3\rho V+O(r^5)\,,\\
\Phi & = & -\frac{1}{8}r^2\rho-r^4\left(\frac{1}{6}T^t{}_t+\frac{1}{64}\rho^2\right)+O(r^6)\,,
\end{eqnarray}
where we remind that $V=V^y$ and $V^x=0$. The change of gauge \eqref{eq:changegauge} implies that we define $R$ asymptotically as
\begin{equation}\label{eq:Rr}
R^{-2}=r^{-2}\left(1-\frac{1}{8}r^2\rho +r^4\left(\frac{1}{12}T^t{}_t-\frac{1}{128}\rho^2+\frac{1}{8}\rho V^2\right)+O(r^6)\right)\,.
\end{equation}
We can invert this order by order to obtain $r=r(R)$ up to any desired power of $R$. Inverting \eqref{eq:Rr} up to order $R^6$ we find
\begin{equation}
r=R\left(1-\frac{1}{16}R^2\rho+\frac{1}{8}R^4\left(\frac{1}{3}T^t{}_t+\frac{1}{64}\rho^2+\frac{1}{2}\rho V^2\right)+O(R^6)\right)\,.
\end{equation}
This can be used to express \eqref{eq:branewithP}--\eqref{eq:branewithP3} with the above expansions for the various functions as an asymptotic solution that is written in terms of the same radial coordinate $R$ as the near-horizon solution. If we carry out these steps we obtain the following expressions for the ansatz functions in the new gauge
\begin{eqnarray}
F_1 & = & 1+\frac{3}{8}R^2\rho+\frac{1}{2}R^4\left(-T^t{}_t+\frac{9}{64}\rho^2-\frac{1}{2}\rho V^2\right)+O(R^6)\,,\label{eq:asympR1}\\
F_2 & = & 1+\frac{3}{8}R^2\rho +R^4\left(-\frac{1}{6}T^t{}_t+\frac{11}{128}\rho^2-\frac{1}{2}\rho V^2\right)+O(R^6)\,,\\
F_3 & = & 1\,,\\
F_4 & = & 1-\frac{1}{4}R^4\rho V^2+O(R^6)\,,\\
N & = & \frac{1}{4}R^3\rho V+O(R^5)\,,\\
G_1 & = & 1+\frac{3}{8}R^2\rho +\frac{1}{2}R^4\left(\frac{9}{64}\rho^2-\frac{1}{4}\rho V^2\right)+O(R^6)\,,\\
G_2 & = & -\frac{1}{4}R^3\rho V+O(R^5)\,,\\
\Phi & = & -\frac{1}{8}R^2\rho-\frac{1}{6}R^4 T^t{}_t+O(R^6)\,.\label{eq:asympR8}
\end{eqnarray}

In order to find an interpolating solution we thus need to solve the equations of motion of \eqref{eq:action3} in the $F_3=1$ gauge such that near the horizon the solution looks like 
\eqref{eq:nearhor1}--\eqref{eq:nearhor8} while near the boundary it looks like \eqref{eq:asympR1}--\eqref{eq:asympR8}. It would be interesting to study the interpolating solution numerically. For the purposes of this work we do not need this explicit solution, but we will need to assume that it exists.

We also see from the asymptotic solution that even for $V=0$ we still can turn on the $\rho$ deformation. Hence static Lifshitz black branes can have a nonzero mass density. Furthermore, even though the full non-linear solution breaks rotational symmetries the near-boundary solution has an asymptotic Killing vector for rotations. Hence rotations are spontaneously broken.

\subsection{Thermodynamics}\label{subsec:thermo}

The most general Killing vector that \eqref{eq:branewithP} admits is of the form 
\begin{equation}
K^M=\left(\partial_t\right)^M+A_1\left(\partial_x\right)^M+A_2\left(\partial_y\right)^M\,,
\end{equation}
where $A_1$ and $A_2$ are constants. The norm is given by
\begin{equation}
\vert\vert K\vert\vert^2=-\frac{F_1}{r^4}+\frac{F_4}{r^2}\left(\frac{N}{r}+A_2\right)^2+A_1^2\frac{F_3}{r^2}\,.
\end{equation}
In order to find the generator of the horizon we demand that $\vert\vert K\vert\vert^2$ vanishes at $R=R_h$ which will be the case if and only if 
\begin{equation}
A_1=0\,,\qquad A_2=-\frac{N(R_h)}{R_h}\,.
\end{equation}
Hence the horizon generator which we will denote by $X^M$ is given by
\begin{equation}\label{eq:horgen}
X^M=\left(\partial_t\right)^M-\frac{N(R_h)}{R_h}\left(\partial_y\right)^M\,.
\end{equation}
We thus see that there is a chemical potential $-N(R_h)/R_h$ associated with the motion in the $y$-direction.

The metric and vector field expanded near the horizon read
\begin{eqnarray}
ds_4^2 & = & -\tilde\rho^2d\tilde t^2+d\tilde\rho^2+\frac{1}{R_h^{2}}dx^2+\frac{p_0}{R_h^2}\left(dy+\frac{N(R_h)}{R_h}dt\right)^2\,,\\
B & = & \frac{1}{2}g_1\left(h_1f_1^{-1}\right)^{1/2}\tilde\rho^2 d\tilde t+\frac{m_0}{R_h}\left(dy+\frac{N(R_h)}{R_h}dt\right)\,,
\end{eqnarray}
where we defined
\begin{eqnarray}
\tilde\rho & = & 2\left(\frac{R-R_h}{h_1 R_h}\right)^{1/2}\,,\\
\tilde t & = & \frac{1}{2}\left(f_1h_1\right)^{1/2}R_h^{-2} t\,.
\end{eqnarray}
We next ask which linear functions $f(t,x,y)$ solve the equation $X^M\partial_M f=0$. These are $x$ and $y+\frac{N(R_h)}{R_h}t$. The metric induced on the common intersection of the hyperplanes $x=\text{cst}$ and $y+\frac{N(R_h)}{R_h}t=\text{cst}$, after Wick rotating the time coordinate $t=-it_E$, is called the bolt and is given by
\begin{equation}
ds^2\vert_{\text{bolt}}=\frac{F_1}{R^4}dt_E^2+\frac{dR^2}{F_2 R^2}\,.
\end{equation}
We expand this metric around $R=R_h$ with a periodic $t_E$ demanding the absence of conical singularities. Because we are on the hyperplane $y+\frac{N(R_h)}{R_h}t=\text{cst}$ this forces us to also Wick rotate $y=-iy_E$ and make it periodic as well in agreement with the interpretation of $-N(R_h)/R_h$ as a chemical potential. The inverse temperature is the periodicity of $t_E$. The temperature and entropy density are given by
\begin{eqnarray}
T & = & \frac{1}{4\pi R_h^2}\left(f_1h_1\right)^{1/2}\,,\label{eq:T}\\
s & = & 4\pi\frac{\left(p_0\right)^{1/2}}{R_h^2}\,,\label{eq:s}
\end{eqnarray}
where we used units in which $16\pi G_N=1$.

In the Wick rotated geometry $t_E$ and $y_E$ are periodic. The thermal cycle parametrized by $t_E$ is contractible while the cycle parametrized by $y_E+\frac{N(R_h)}{R_h}t_E$ is non-contractible. Hence we can compute $\oint_{R=R_h}B$ where we integrate along the cycle parametrized by $y_E+\frac{N(R_h)}{R_h}t_E$. The result is
\begin{equation}
\oint_{R=R_h}B=\frac{4\pi m_0 n_0}{\left(f_1 h_1\right)^{1/2}}\,.
\end{equation}
In general using our ansatz the massive vector field can be written as
\begin{equation}
B^M=-R^2\frac{G_1}{F_1}\left(\left(\partial_t\right)^M-\frac{N}{R}\left(\partial_y\right)^M\right)+R\frac{G_2}{F_4}\left(\partial_y\right)^M\,.
\end{equation}
It thus follows that for $m_0=0$ the massive vector field $B^M$ is proportional to the horizon generator $X^M$ at $R=R_h$.

It can be shown by using the near-horizon solution that the charges $Q_\lambda$ and $Q_\mu$ \eqref{eq:Qlambda} and \eqref{eq:Qnu} are such that
\begin{equation}\label{eq:thermorel}
Q_\lambda-\frac{3}{2}Q_\mu=Ts\, , 
\end{equation}
with $T$, $s$ given in \eqref{eq:T}, \eqref{eq:s}. 
Using the asymptotic form of the solution \eqref{eq:asymp1}--\eqref{eq:asymp6} with \eqref{eq:bdryPF1} and \eqref{eq:bdryPF2} to compute the left hand side of \eqref{eq:thermorel} we conclude that
\begin{equation}\label{eq:thermorel2}
\mathcal{E}+P=Ts+\frac{1}{2}\rho V^2\,.
\end{equation}
The equations of state follows from \eqref{eq:z=2trace}
\begin{equation}
P=\mathcal{E}\,.
\end{equation}

We have thus been able to derive the thermodynamic relations without knowing the full solution analytically using the Noether charges $Q_\lambda$ and $Q_\mu$. This is similar to what has been done in \cite{Bertoldi:2009vn,Bertoldi:2009dt}. We will see further below that we can also derive the first law of thermodynamics without having full analytic control of the solution. All that we need to know is the near-horizon expansion, the near-boundary expansion and the existence of an interpolating solution. We assume the latter to be the case. It would be interesting to provide numerical evidence for the interpolating solution.

\subsection{Charges}\label{subsec:charges}

The goal of this subsection is to find an alternative derivation of \eqref{eq:thermorel2} which can be thought of as an integral form in terms of the renormalized on-shell action and certain horizon charges. The second goal is to find additional relations between near-boundary and near-horizon quantities. In particular we will show that the velocity $V^y=V$ is equal to the chemical potential $-N(R_h)/R_h$.

In order to define the black brane charges we use the boundary diffeomorphism Ward identity which on a flat TNC geometry reads \eqref{eq:WILifhydro2}. Given a boundary Killing vector $K^\mu$ in the sense that
\begin{equation}
\mathcal{L}_K\tau_\mu=0\,,\qquad\mathcal{L}_K\bar h_{\mu\nu}=0\,,\qquad\mathcal{L}_K\tilde\Phi=0\,,
\end{equation}
it can be shown (see \cite{Hartong:2014pma,Hartong:2015wxa}) that we find the conserved current
\begin{equation}
\partial_\mu \left(K^\nu T^\mu{}_\nu\right)=0\,.
\end{equation}
The conserved charge associated with the boundary Killing vector $K^\mu$ is thus
\begin{equation}\label{eq:QK}
Q_K=-\int_{t=\text{cst}}dxdy K^\nu T^t{}_\nu\,.
\end{equation}
For our case the integrand is independent of $x$ and $y$ and so it is better to consider the charge per unit boundary volume. We will often write $\int_{t=\text{cst}}dxdy$ as a formal integral that we never really perform. We can always divide the charges by it. We will assume that $K^\mu$ is the $\mu$ component of a bulk Killing vector $K^M$.

Using the definitions of the vevs in \eqref{eq:S0}, \eqref{eq:Sa} and \eqref{eq:vev1}, \eqref{eq:vev2}{}\footnote{The quantity $\alpha_{(0)}$ defined in \eqref{eq:alpha0} equals unity because for our solutions $\Phi=O(r^2)$ so that $\phi=0$ as follows from \eqref{eq:sources4}.}, as well as the boundary energy-momentum tensor in \eqref{eq:chiT} and \eqref{eq:TandTchi} we find that
\begin{equation}
T^t{}_\nu=-\lim_{r\to 0} r^{-2}\left(T_{\mu\nu}E^\nu_0+\mathcal{T}^\rho E^0_\rho B_\mu\right)\,.
\end{equation}
Using \eqref{eq:Tmunu} and \eqref{eq:calT} we find that for purely radial solutions (no dependence on boundary coordinates)
\begin{equation}
T_{\mu\nu}E^\nu_0+\mathcal{T}^\rho E^0_\rho B_\mu=\frac{1}{\sqrt{-h}}\mathcal{L}_{\text{bdry}}^{\text{os}}E_\mu^0+2K_{\mu\nu}E^\nu_0+e^{3\Phi}n^M E^\nu_0 F_{M\nu}B_\mu\,,
\end{equation}
where $\mathcal{L}_{\text{bdry}}^{\text{os}}$ is the on-shell value of the counterterm Lagrangian \eqref{eq:counterterms4D} including the Gibbons--Hawking boundary term, i.e.
\begin{equation}
\mathcal{L}_{\text{bdry}}^{\text{os}}=\sqrt{-h}\left(2K-5e^{-\Phi/2}+e^{\Phi/2}B_\rho B^\rho\right)\,.
\end{equation}
The extrinsic curvature $K$ is given by $K=h^{\mu\nu}K_{\mu\nu}$ where $K_{\mu\nu}$ is the $\mu\nu$ component of $K_{MN}=-\frac{1}{2}\mathcal{L}_n h_{MN}=\nabla_M n_N-n_M n^K\nabla_K n_N$ with the unit normal vector $n_M$ given by $n_M=-(g^{rr})^{-1/2}\delta_M^r$. Since the Killing vector $K^M$ is a boundary Killing vector we have $K^M n_M=0$. Further we employ a radial gauge choice such that $E^M_0 n_M=0$. Using these results we can write 
\begin{equation}
Q_K=\int_{t=\text{cst}}dxdy \lim_{r\to 0} r^{-2}\left(\frac{1}{\sqrt{-h}}\mathcal{L}_{\text{bdry}}^{\text{os}}K^ME_M^0+n^M E^{N0} Z_{NM}\right)\,,
\end{equation}
where $Z_{NM}=-Z_{MN}${}\footnote{The antisymmetry follows from the fact that $K_M$ is also assumed to be a bulk Killing vector.} is given by
\begin{equation}\label{eq:tildeZ}
Z_{NM}=2\nabla_N K_M+e^{3\Phi}F_{NM}K^PB_P\,.
\end{equation}

The integrand is over a $t=\text{cst}$ hypersurface. Its timelike unit normal is given by 
\begin{equation}
u_M=U\delta_M^t\,,\qquad U=r^{-2}\left(1+\frac{1}{16}r^2\rho +r^4\left(-\frac{1}{6}T^t{}_t+\frac{5}{2}\frac{1}{256}\rho^2\right)+O(r^6)\right)\,,
\end{equation}
where we used the boundary expansions of section \ref{subsec:asympsol}. It can be shown using these same expansions that
\begin{equation}
U^t=E^{t0}+O(r^8)\,,\qquad U^i=O(r^4)\,,
\end{equation}
where $E^{t0}=U^{-1}+O(r^8)$. We also have that $E^{i0}=O(r^4)$. Using these results together with the near-boundary expansion of $Z_{NM}$ it can be proven that we can replace $E^{N0}$ by $u^N$ everywhere in the integrand of $Q_K$, i.e. we can write 
\begin{equation}
Q_K=\int_{t=\text{cst}}dxdy \lim_{r\to 0} r^{-2}\left(\frac{1}{\sqrt{-h}}\mathcal{L}_{\text{bdry}}^{\text{os}}K^Mu_M+n^M u^NZ_{NM}\right)\,.
\end{equation}

Let us define the projector $P_M^N=\delta_M^N+u_M u^N$ which projects onto the $t=\text{cst}$ hypersurface whose metric we will denote by $H_{IJ}$, i.e. using the ADM decomposition we obtain
\begin{equation}
ds^2=-U^2dt^2+H_{IJ}\left(dx^I+u^Idt\right)\left(dx^J+u^Jdt\right)\,.
\end{equation}
Let us furthermore define $Z^M=u^N Z_{NM}$. We can derive the following identity
\begin{equation}
P_M^N\nabla_N Z^M=\frac{1}{\sqrt{-H}}\partial_I\left(\sqrt{H}Z^I\right)\,.
\end{equation}
Hence it follows that
\begin{equation}
\int_{t=\text{cst}}d^3x\sqrt{H}P_M^N\nabla_N Z^M=\int_{t=\text{cst}}dxdy\int_\epsilon^{R_h}dR\partial_R\left(\sqrt{H}Z^R\right)\,,
\end{equation}
where we used the radial $R$ coordinate of section \ref{subsec:interpolating}, i.e. the $F_3=1$ gauge, with a cut-off boundary at $R=\epsilon$ and the horizon at $R=R_h$. The integration measure in terms of the ansatz functions can be written as
\begin{equation}
\sqrt{H}=r^{-3}F_{2}^{-1/2}F_4^{1/2}\,.
\end{equation}
It follows that 
\begin{eqnarray}
\int_{t=\text{cst}}d^3x\sqrt{H}P_M^N\nabla_N Z^M & = & \int_{t=\text{cst}}dxdy R^{-2}n_M Z^M\vert_{R=\epsilon}\nonumber\\
&&+\int_{t=\text{cst}}dxdy R^{-3}F_{2}^{-1/2}F_4^{1/2}Z^R\vert_{R=R_h}\,.
\end{eqnarray}
We conclude that the charge $Q_K$ can be written as
\begin{eqnarray}
Q_K & = & \int_{t=\text{cst}}dxdy R^{-2}\frac{1}{\sqrt{-h}}\mathcal{L}_{\text{bdry}}^{\text{os}}K^MU_M\vert_{R=\epsilon}+\int_{t=\text{cst}}dxdy\int_\epsilon^{R_h}dR\sqrt{H}P_M^N\nabla_N Z^M\nonumber\\
&&+\int_{t=\text{cst}}dxdy R^{-2}F_4^{1/2}X^NY^P  Z_{NP}\vert_{R=R_h}\,,
\end{eqnarray}
where we send $\epsilon$ to zero. In the horizon integral $X^P$ is the horizon generator \eqref{eq:horgen} and $Y^N$ is given by (in the EF coordinates of \eqref{eq:ansatzEF1} with $F_3=1$)
\begin{equation}
Y^N=R^3\left(\frac{F_2}{F_1}\right)^{1/2}\delta^N_R\,.
\end{equation}
The vector $Y$ is a null vector that satisfies $X\cdot Y=-1$ at the horizon.

Using the equations of motion \eqref{eq:Einsteineq}--\eqref{eq:vectoreom} with \eqref{eq:upliftablemodel} as well as the fact that the Killing vector $K$ is a symmetry of the matter fields which means that
\begin{equation}
\mathcal{L}_K B_M=0\,,\qquad\mathcal{L}_K\Phi=0\,,
\end{equation}
it can be shown that
\begin{equation}
P_M^N\nabla_N Z^M=\frac{1}{\sqrt{-g}}\mathcal{L}_{\text{bulk}}^{\text{os}}K^Mu_M\,.
\end{equation}
The charge can now be written as
\begin{equation}\label{eq:charge}
Q_K=K^tTS_E^{\text{os}}+\int_{t=\text{cst}}d^2x\sqrt{\sigma} X^NY^P  Z_{NP}\vert_{R=R_h}\,,
\end{equation}
where we used that $K^Mu_M=K^tU$ with $\sqrt{-g}=U\sqrt{H}$ and where we defined $\sqrt{\sigma}=F_4^{1/2}R^{-2}$ which is the determinant of the metric on the $t=\text{cst}$ and $R=\text{cst}$ submanifold. In this expression for the charge $S_E^{\text{os}}$ is the Euclidean on-shell action, i.e.
\begin{equation}
TS_E^{\text{os}}=\lim_{\epsilon\to 0}\left[\int_\epsilon^{R_h}dR\int_{t=\text{cst}}d^2x\mathcal{L}_{\text{bulk}}^{\text{os}}+\int_{t=\text{cst}}d^2x \mathcal{L}_{\text{bdry}}^{\text{os}}\vert_{R=\epsilon}\right]\,.
\end{equation}
Equation \eqref{eq:charge} is the result we were looking for. It expresses the asymptotic charge associated with the Killing vector $K^M$ in terms of a horizon integral and the Euclidean on-shell action.

This result can be used to compute the charges associated with the Killing vectors $\partial_t$ and $\partial_y$ twice, once near the boundary using \eqref{eq:QK} and once at the horizon using \eqref{eq:charge}. Near the boundary we find
\begin{eqnarray}
Q_{\partial_t} & = & -\int_{t=\text{cst}}dxdy T^t{}_t=\int_{t=\text{cst}}dxdy\left(\mathcal{E}+\frac{1}{2}\rho V^2\right)\,,\\
Q_{\partial_y} & = & -\int_{t=\text{cst}}dxdy T^t{}_y=-\int_{t=\text{cst}}dxdy\rho V\,.\label{eq:Qy}
\end{eqnarray}
Using \eqref{eq:charge} we can derive the following relation
\begin{equation}\label{eq:quanstatrel}
Q_{\partial_t}-TS_E^{\text{os}}=T\int_{t=\text{cst}}dxdy s-\frac{N(R_h)}{R_h}Q_{\partial_y}\,,
\end{equation}
where we used \eqref{eq:T} and \eqref{eq:s}. 

The momentum $Q_{\partial_y}$ can be written in terms the Noether charge $Q_\mu$ defined in \eqref{eq:Qnu} via 
\begin{equation}
\frac{N(R_h)}{R_h}Q_{\partial_y}=\int_{t=\text{cst}}dxdy Q_\mu\,.
\end{equation}
This can be proven by computing the left and right hand side at the horizon where for the left hand side we use the integral form given in \eqref{eq:charge}. The Noether charge $Q_\mu$ can also be computed near the boundary where it gives $Q_\mu=\rho V^2$. Hence with \eqref{eq:Qy} we conclude that 
\begin{equation}
\frac{N(R_h)}{R_h}=-V^y=-V\,,
\end{equation}
i.e. the chemical potential is the velocity of the fluid. From this and \eqref{eq:quanstatrel} it follows that the Euclidean on-shell action relates to the pressure as follows
\begin{equation}\label{eq:Euclidosaction}
TS_E^{\text{os}}=- \int_{t=\text{cst}}dxdy P\,.
\end{equation}

We believe that similar arguments allow one to derive the first law of thermodynamics for these Lifshitz holographic fluids. For example using arguments similar to those of \cite{Papadimitriou:2005ii} that do not require an explicit knowledge of the interpolating solution. However there are quite compelling arguments that fix the first law in a more straightforward manner so we will refrain from using a more general approach. One of these arguments uses the Ward identities of the dual holographic fluid and the existence of an entropy current. This will be discussed in the next section. The other argument results from the assumption that the pressure only depends on temperature and chemical potential. Given, say, a numerical solution this could be tested by evaluating \eqref{eq:Euclidosaction}. For us this is a rather minor assumption because it is essentially assuming that a solution with a horizon generated by $X^M$ exists. If we assume that $P=P(T, V^2)$ we can vary it and use \eqref{eq:thermorel2} to derive
\begin{equation}
\left(\frac{\partial P}{\partial T}\right)_{V^2}=s\,,\qquad \left(\frac{\partial P}{\partial V^2}\right)_{T}=\frac{1}{2}\rho\,,\qquad\delta\mathcal{E}=T\delta s+\frac{1}{2}V^2\delta\rho\,,
\end{equation}
where the latter relation is the first law for our holographic Lifshitz perfect fluid. More will be said about this in the next section.

\section{Lifshitz perfect fluids}\label{subsec:WIs}

This section is independent from holography and derives the Lifshitz perfect fluid from dimensional reduction. In appendix \ref{subsec:nullredfluid} we have discussed the null reduction of a relativistic perfect fluid. This gives rise to a Galilean perfect fluid. If furthermore the relativistic fluid is scale invariant, i.e. conformal, the lower-dimensional Galilean perfect fluid has a $z=2$ Schr\"odinger invariance. The $z=2$ Schr\"odinger algebra contains the $z=2$ Lifshitz algebra as a subalgebra. Hence a Lifshitz invariant system can be obtained by starting with a Schr\"odinger invariant system and breaking the generators that are part of the Schr\"odinger algebra but not of the Lifshitz algebra. One of these symmetries is particle number $N$. By breaking $N$ explicitly the $z=2$ Schr\"odinger algebra reduces to the $z=2$ Lifshitz algebra\footnote{In the Schr\"odinger algebra the commutator between Galilean boosts $G_i$ and momenta $P_i$ reads $[P_i\,,G_j]=\delta_{ij}N$ so by breaking $N$ keeping $P_i$ intact we break $G_i$. Further special conformal symmetries $K$ in the Schr\"odinger algebra satisfy the commutation relation $[K\,,P_i]=-G_i$ so that breaking $G_i$ leads to broken $K$ symmetries. Hence by breaking $N$ we loose the $G_i$ and $K$ generators as well and we are left with the Lifshitz algebra.}. This is precisely what our holographic model for Lifshitz invariant field theories does. 

We have shown that the 4-dimensional bulk theory follows from Scherk--Schwarz reduction of a 5-dimensional AdS-gravity model coupled to a scalar field. This scalar field leads to an additional source in the dual field theory and, as derived in section \ref{subsec:onepointfunctions}, the corresponding diffeomorphism Ward identity,  reads 
\begin{equation}\label{eq:nullreductionwithchi1}
\nabla_{A}t^{A}{}_{B}=-\langle O_\psi\rangle\partial_{B}\psi\,.
\end{equation}
Here we will be interested in flat space only so the left hand side is simply $\nabla_{A}t^{A}{}_{B}=\partial_A t^A{}_B$. The Scherk--Schwarz reduction tells us that 
\begin{eqnarray}
\psi & = & 2u-2\chi\,,\label{eq:nullreductionwithchi2}\\
\langle O_\psi\rangle & = & -\frac{1}{2}\langle O_\chi\rangle\,,\label{eq:nullreductionwithchi3}
\end{eqnarray} 
where $\chi$ and $\langle O_\chi\rangle$ are independent of $u$. If we now set the 4-dimensional scalar source $\chi=0$ we obtain the 4-dimensional Ward identities (see also \eqref{eq:WILifhydro2} and \eqref{eq:WILifhydro3})
\begin{eqnarray}
\partial_\mu T^\mu{}_\nu & = & 0\,,\label{eq:LifWI1}\\
\partial_\mu T^\mu & = & \langle O_\chi\rangle\,,\label{eq:LifWI2}
\end{eqnarray}
where we used \eqref{eq:nullredt6} and \eqref{eq:nullredt7}. We thus see that the mass current $T^\mu$ is not conserved due to the presence of $\langle O_\chi\rangle$. The $z=2$ scale Ward identity follows from \eqref{eq:nullredtraceWI} with $t^A{}_A=0$ which for the case of a flat TNC space-time \eqref{eq:flatTNC} reads
\begin{equation}\label{eq:LifWI3}
2T^t{}_t+T^i{}_i=0\,.
\end{equation}
The null reduction also implies the identities \eqref{eq:boostWI} and \eqref{eq:rotationWI} which on a flat TNC space-time read
\begin{equation}\label{eq:LifWI4}
T^t{}_i  = T^i\,,\qquad T^i{}_j=T^j{}_i\,.
\end{equation}

The null reduction in the presence of the scalar source $\psi$ as written in \eqref{eq:nullreductionwithchi1}--\eqref{eq:nullreductionwithchi3} gives rise to a system that breaks Galilean boost symmetries and particle number. This is due to the fact the $\psi$ in \eqref{eq:nullreductionwithchi2} breaks these symmetries. What we are left with is a $z=2$ Lifshitz invariant system in one dimension lower. 

We will now apply the Lifshitz Ward identities \eqref{eq:LifWI1}--\eqref{eq:LifWI4} to the case of a $d=z=2$ perfect fluid where $\mathcal{E}=P$ and $V^i$ are functions of $t,x^i$. It has been shown that the form of $T^\mu{}_\nu$ and $T^\mu$ for the null reduction of a relativistic perfect fluid take the form \eqref{eq:PFEMT3} and \eqref{eq:PFEMT4}. We now consider the fluid equations as follows from the Ward identities and demand that there exists a conserved entropy current. The latter requirement will tell us what the thermodynamic relations for a Lifshitz perfect fluid are.

On flat TNC space-time the form of the fluid energy-momentum tensor and mass current for a perfect fluid are given by \eqref{eq:bdryPF1} and \eqref{eq:bdryPF2}. The fluid equations are thus given by the Ward identities which read\footnote{If in the holographic setup we would make the fluid variables functions of the boundary coordinates we would have to correct the energy-momentum tensor by derivatives of the fluid variables. The Einstein equations will then lead to Ward identities for  this corrected boundary energy-momentum tensor. At leading order in derivatives it will however reduce to the Ward identities for a perfect fluid.}
\begin{eqnarray}
0 & = & \partial_t\left(\mathcal{E}+\frac{1}{2}\rho V^2\right)+\partial_i\left(\left(\mathcal{E}+P+\frac{1}{2}\rho V^2\right)V^i\right)\,,\label{eq:Econserv}\\
0 & = & \partial_t\left(\rho V_i\right)+\partial_i\left(P\delta^j_i+\rho V^j V_i\right)\,,\\
\langle O_\chi\rangle & = & \partial_t\rho+\partial_i\left(\rho V_i\right)\,.\label{eq:massnonconserv}
\end{eqnarray}
These equations can be used to rewrite the equation for energy conservation \eqref{eq:Econserv} as
\begin{equation}
\partial_t\mathcal{E}+V^i\partial_i\mathcal{E}+\left(\mathcal{E}+P-\frac{1}{2}\rho V^2\right)\partial_i V^i-\frac{1}{2}V^2\left(\partial_t\rho+V^i\partial_i\rho\right)=0\,.
\end{equation}
This gives rise to an equation for conservation of entropy,
\begin{equation}
\partial_t s+\partial_i\left(sV^i\right)=0\,,
\end{equation}
provided we take
\begin{eqnarray}
\mathcal{E}+P & = & Ts+\frac{1}{2}\rho V^2\,,\\
\delta\mathcal{E} & = & T\delta s+\frac{1}{2}V^2\delta\rho\,.\label{eq:firstlaw}
\end{eqnarray}

These two equations together with the equation of state $P=\mathcal{E}$ (which follows from \eqref{eq:LifWI3}) describe the thermodynamic properties of a Lifshitz invariant system obtained by breaking particle number symmetries. What we see here is a realization of a Lifshitz perfect fluid where the velocity or rather, due to rotational symmetries, $V^2$, plays the role of a chemical potential.%
\footnote{A similar extension of the first law of thermodynamics involving a fluid with
boost momentum  was seen in \cite{Armas:2015nea} in the proposed effective theory for the dynamics of
helicoidal black $p$-branes using the blackfold construction \cite{Emparan:2009cs,Emparan:2009at}.}
 The thermodynamically conjugate variable is the mass density $\rho$. From the first law \eqref{eq:firstlaw} it follows that 
\begin{equation}
\delta P = s\delta T+\frac{1}{2}\rho \delta V^2\,,
\end{equation}
so that pressure is a function of $T$ and the chemical potential $V^2$.

We see here that the way in which we realize Lifshitz hydrodynamics is quite different from what has been discussed in \cite{Hoyos:2013eza}. The approach in \cite{Hoyos:2013eza} is to start with a $z=1$ relativistic perfect fluid and to break Lorentz symmetries by adding higher derivative interactions that break the symmetry of the energy-momentum tensor. One can then take a non-relativistic limit to obtain systems with $z\neq 1$ that break Galilean boost symmetries. This leads to a model where Galilean boosts are broken at higher orders in a derivative expansion. On the other hand here we realize Lifshitz symmetries by breaking particle number and hence Galilean boosts already at the perfect fluid level. In \cite{Hartong} we will present more examples of Lifshitz hydrodynamics from a field theory perspective.

As a final comment we note that in order to solve the $d+2$ equations \eqref{eq:Econserv}--\eqref{eq:massnonconserv} we need to know what $\langle O_\chi\rangle$ is in terms of the fluid variables $\rho$, $V^i$ and $\mathcal{E}$. Explicit examples will be given in \cite{Hartong}.

\section{Discussion and Outlook \label{sec:out} }

We have shown that there is a new class of Lifshitz perfect fluids in which Galilean boosts are broken at the perfect fluid level. The holographic dual description is realized by a moving black brane solution of the EPD model. The motion of the black brane is not obtained by applying a boost transformation to a static black brane but follows from constructing a new class of solutions corresponding to Lifshitz black branes with linear momentum. From the dual field theory point of view the boundary fluid can be obtained by a twisted null reduction of relativistic fluid in the background of a free scalar source that depends linearly on the null circle. From the lower-dimensional point of view this corresponds to a Schr\"odinger fluid with broken particle number symmetry.

In this work we restricted our attention to a specific EPD model for which we obtained the counterterms and near-boundary expansion by dimensional reduction from AdS holography coupled to a free real scalar. In order to consider similar solutions of other EPD models we need to be able to write down the counterterms and near-boundary expansions for general EPD models. Despite a lot of effort the situation is presently still not fully understood. There are different proposals \cite{Chemissany:2014xpa,Chemissany:2014xsa} and \cite{Hartong:2014oma,Hartong:2015wxa} (see  \cite{Ross:2011gu,Baggio:2011cp} for earlier work) that share certain similarities but that also have some differences. A comparison between \cite{Chemissany:2014xpa,Chemissany:2014xsa} and \cite{Hartong:2014oma,Hartong:2015wxa} is made in appendix \ref{subsec:comparison}. We believe that more work needs to be done before we can state what the near-boundary expansion and counterterms are for a given EPD model in the general class that admits Lifshitz solutions. This general analysis includes asymptotically Lifshitz solutions with hyperscaling violation exponent $\theta$ and the charge hyperscaling violation exponent introduced in \cite{Gath:2012pg,Gouteraux:2012yr}.

A special subset of the EPD models are those for which $W=0$ so that the bulk vector field becomes a Maxwell gauge potential with a $U(1)$ gauge symmetry. It has been shown in \cite{Kiritsis:2015doa} that the corresponding global $U(1)$ symmetry in the boundary theory leads to mass conservation. For the EMD model we know the black brane solutions that are dual to perfect fluids analytically \cite{Taylor:2008tg,Charmousis:2010zz}. For the solutions of the EPD models with $W\neq 0$ we only know the solution near the boundary and near the horizon but we do not know that interpolating solution. Hence we have to resort to arguments based on the existence of conserved Noether charges that are a consequence of various ansatz symmetries that allows one to relate near-boundary and near-horizon properties of the solution as was done in \cite{Bertoldi:2009vn,Bertoldi:2009dt}. Here we followed a similar approach and we added to this various integral forms of the asymptotic charges related to the existence of Killing vectors. It would be interesting to see how far one can push this kind of analysis beyond the perfect fluid level. In other words it is worth exploring if it possible to construct bulk solutions in which the fluid variables such as the temperature and velocity become slowly varying functions of the boundary coordinates in such a way that we can extract all the relevant boundary properties from the near-horizon and near-boundary features. Further, it would be nice to have numerical confirmation about the interpolating solution we have assumed to exist. 

We also remark that it would be interesting to study the role of charge in the boundary Lifshitz hydrodynamics by adding additional $U(1)$ vector gauge fields to the bulk description like in \cite{Tarrio:2011de}. For this the recent results in \cite{Festuccia2016b} on non-relativistic electrodynamics coupled to TNC could be relevant.

Finally, another interesting direction to pursue is to  use Ho\v rava--Lifshitz gravity theories as bulk theories in holography
 \cite{Griffin:2012qx,Janiszewski:2012nf} and examine the connection with Lifshitz hydrodynamics \cite{Eling:2014saa,Davison:2016auk}. It would be  worthwhile to pursue this further in the light of the results of this paper. In particular in connection to dynamical NC geometry \cite{Hartong:2015zia} and  finite temperature states in the 3-dimensional Chern-Simons Schr\"odinger gravity that was recently constructed \cite{Hartong:2016yrf}.

\section*{Acknowledgments}

We would like to thank Jay Armas, Jan de Boer, Sa\v so Grozdanov, Kristan Jensen, Cynthia Keeler, Elias Kiritsis, Jorge Santos, Koenraad Schalm, Watse Sybesma, Javier Tarrio and Stefan Vandoren for valuable discussions. The work of JH is partially supported by a Marina Solvay fellowship as well as by the advanced ERC grant `Symmetries and Dualities in Gravity and M-theory' of Marc Henneaux. The work of NO and MS is supported in part by the Danish National Research Foundation project  ``New horizons in particle and condensed matter physics from black holes".

\appendix

\section{Torsional Newton--Cartan geometry and non-relativistic field theory}\label{app:TNC}

Here we summarize results obtained in \cite{Andringa:2010it,Christensen:2013lma,Christensen:2013rfa,Jensen:2014aia,Hartong:2014oma,Hartong:2014pma,Bergshoeff:2014uea,Hartong:2015wxa,Hartong:2015zia} regarding the most general formulation of torsional Newton--Cartan (TNC) geometry. We will focus only on those aspects that are needed for the purposes of this work. Since here we encounter TNC geometry through null reduction of the AdS$_5$ boundary metric we will study its properties in this context.

\subsection{Null reduction of metric}\label{subsec:nullred}

Consider the null reduction ansatz for the metric
\begin{equation}\label{eq:higherDmetric}
ds^2=\gamma_{AB}dx^{A}dx^{B}=2\tau_\mu dx^\mu\left(du-m_\nu dx^\nu\right)+h_{\mu\nu}dx^\mu dx^\nu\,,
\end{equation}
where $A=(u,\mu)$ and with
\begin{equation}\label{eq:h}
h_{\mu\nu}=\delta_{ab}e_\mu^a e^\nu_b\,,
\end{equation}
in which $a=1\,,\ldots,d$. The metric $h_{\mu\nu}$ has vanishing determinant. The reduction ansatz is the most general metric for which $\gamma_{uu}=0$. It is assumed that $\partial_u$ is a null Killing vector of $\gamma_{AB}$. The fields $\tau_\mu$ and $e^a_\mu$ are the vielbeins of the $d+1$ dimensional TNC geometry. The metric \eqref{eq:higherDmetric} preserves the following local tangent space transformations 
\begin{eqnarray}
\delta \tau_\mu & = & 0 \,, \\
\delta e_\mu^a & = & \tau_\mu\lambda^a + \lambda^{a}{}_b e_{\mu}^{b}  \,,\\
\delta m_\mu & = & \partial_\mu\sigma+ \lambda_a e_{\mu}^a\,.
\end{eqnarray}
The local $\sigma$ transformation requires $\delta u=\sigma$. The transformations with local parameter $\lambda^a$ correspond to tangent space Galilean boosts $(G)$ and transformations with local parameter $\lambda^a{}_b$ correspond to tangent space rotations $(J)$. The metric components $\gamma_{\mu u}=\tau_\mu$ and $\gamma_{\mu\nu}=\bar h_{\mu\nu}$ where 
\begin{equation}\label{eq:barh}
\bar h_{\mu\nu}=h_{\mu\nu}-\tau_\mu m_\nu-\partial_\nu\tau_\mu\,,
\end{equation}
are invariant under these local transformations.

The inverse metric is
\begin{equation}
\gamma^{uu}=2\tilde\Phi\,,\qquad \gamma^{u\mu}=-\hat v^\mu\,,\qquad \gamma^{\mu\nu}=h^{\mu\nu}\,,
\end{equation}
where 
\begin{eqnarray}
\tilde\Phi & = & -v^\rho m_\rho+\frac{1}{2}h^{\rho\sigma}m_\rho m_\sigma\,,\label{eq:tildePhi}\\
\hat v^\mu & = & v^\mu-h^{\mu\nu}m_\nu\,,\label{eq:hatv}\\
h^{\mu\nu} & = & \delta^{ab}e^\mu_a e^\nu_b\,.
\end{eqnarray}
The inverse vielbeins $v^\mu$ and $e^\mu_a$ are defined through
\begin{eqnarray}
v^\mu e_\mu^a & = & 0\,,\\
v^\mu\tau_\mu & = & -1\,,\\
e^{\mu}_a\tau_\mu & = & 0\,,\\
e^{\mu}_ae_{\mu}^b & = & \delta^b_a\,.
\end{eqnarray}
It is sometimes useful to work with the Galilean boost invariant vielbeins $\tau_\mu$, $\hat e_\mu^a$ and their inverse $\hat v^\mu$, $e^\mu_a$ where $\hat e_\mu^a$ is defined by
\begin{equation}\label{eq:hate}
\hat e_\mu^a = e_\mu^a-m_\nu e^{\nu a}\tau_\mu\,.
\end{equation}
These satisfy the orthogonality relations 
\begin{eqnarray}
\hat v^\mu\hat e_\mu^a & = & 0\,,\\
\hat v^\mu\tau_\mu & = & -1\,,\\
e^{\mu}_a\tau_\mu & = & 0\,,\\
e^{\mu}_a\hat e_{\mu}^b & = & \delta^b_a\,.
\end{eqnarray}
Finally we will often use the spatial metric $\hat h_{\mu\nu}$ defined by
\begin{equation}
\hat h_{\mu\nu}=\delta_{ab}\hat e_\mu^a\hat e_\nu^b=\bar h_{\mu\nu}+2\tilde\Phi\tau_\mu\tau_\nu\,.
\end{equation}
The inverse vielbeins transform as
\begin{eqnarray}
\delta v^\mu & = & \lambda^a e^\mu_a \,, \\
\delta e^\mu_a & = &  \lambda_{a}{}^b e^{\mu}_{b}  \,.
\end{eqnarray}

A torsionful affine connection $\Gamma^{\rho}_{\mu\nu}$ that is invariant under the local tangent space symmetries $(G,J)$ and that satisfies metric compatibility, in the sense of 
\begin{eqnarray}
\nabla_\mu\tau_\nu & = & 0\,,\label{eq:TNC1}\\
\nabla_\mu h^{\nu\rho} & = & 0\,,
\end{eqnarray}
is given by 
\begin{eqnarray}
\bar\Gamma^{\rho}_{\mu\nu} & = & -\hat v^\rho\partial_\mu\tau_\nu+\frac{1}{2}h^{\rho\sigma}\left(\partial_\mu\bar h_{\nu\sigma}+\partial_\nu \bar h_{\mu\sigma}-\partial_\sigma\bar h_{\mu\nu}\right)\,.\label{eq:TNCconnection}
\end{eqnarray}
The bar on $\Gamma$ is supposed to emphasize that the connection is not unique and we have chosen a particular realization. We will later encounter another affine connection that is metric compatible.  We note that in  \cite{Festuccia2016a} it is shown that at the linearized level the connection \eqref{eq:TNCconnection}
(which is linear in $m_\mu$) appears when applying the Noether procedure to gauging the space-time symmetries in theories with Galilean symmetries.

\subsection{Null reduction of energy-momentum tensor}\label{subsec:nullredEMT}

In \cite{Christensen:2013rfa,Jensen:2014aia,Hartong:2014oma,Hartong:2014pma,Hartong:2015wxa} (see also \cite{Festuccia2016a}) the coupling prescriptions of non-relativistic field theories to torsional Newton--Cartan (TNC) backgrounds have been worked out both directly in field theory and from Lifshitz holography. The results of course agree (see e.g. \cite{Hartong:2015wxa}). Here we briefly review these results and derive them from null reduction as done in \cite{Christensen:2013rfa}.

The TNC energy-momentum tensor (EMT) is defined as the response to varying the TNC fields via
\begin{eqnarray}
\delta_{\text{bg}} S & = & \int d^{d+1}x e\left[-\tau_\nu T^\nu{}_\mu\delta\hat v^\mu-\left(\hat h_{\sigma\nu}\hat v^\mu T^\nu{}_\mu\right)\tau_\rho\delta h^{\rho\sigma}\right.\nonumber\\
&&\left.+\frac{1}{2}\left(\hat h_{\rho\nu}\hat h_{\sigma\lambda}h^{\lambda\mu}T^\nu{}_\mu\right)\delta h^{\rho\sigma}+\tau_\mu T^\mu\delta\tilde\Phi\right]\,,\label{eq:variationTNC}
\end{eqnarray}
where $e$ is the determinant of the 3 by 3 matrix $(\tau_\mu, e^a_\mu)$ which is both boost and rotation invariant. We can alternatively define an energy-momentum tensor by varying the unhatted TNC fields via
\begin{eqnarray}
\delta_{\text{bg}} S & = & \int d^{d+1}x e\left[-\mathcal{T}_\mu\delta v^\mu+\frac{1}{2}\mathcal{T}_{\mu\nu}\delta h^{\mu\nu}+T^\mu\delta m_\mu\right]\,.\label{eq:variation4}
\end{eqnarray}
The two are related via
\begin{equation}\label{eq:calS}
h^{\nu\rho}\mathcal{T}_{\rho\mu}-v^\nu\mathcal{T}_\mu=T^\nu{}_\mu+T^\nu m_\mu\,.
\end{equation}

According to the null reduction of \cite{Christensen:2013rfa} the energy momentum tensor $T^\mu{}_\nu$ and mass current $T^\mu$ are related to the 
higher-dimensional energy-momentum tensor $t^{AB}$ via\footnote{Since $t^{AB}$ is the response to varying $\gamma_{AB}$ there is no need for $t^{uu}$ since 
$\gamma_{uu}=0$.}
\begin{eqnarray}
t^{\mu u} & = & 2\tilde\Phi T^\mu-\hat v^\sigma T^\mu{}_\sigma\,,\label{eq:invtumu}\\
t^{\mu\nu} & = & -\hat v^\mu T^\nu+h^{\mu\rho}T^\nu{}_\rho\,.\label{eq:invtmunu}
\end{eqnarray}
The latter relation implies due to the symmetry of $t^{\mu\nu}$
\begin{equation}
-\hat v^\mu T^\nu+h^{\mu\rho}T^\nu{}_\rho+\hat v^\nu T^\mu-h^{\nu\rho}T^\mu{}_\rho=0\,,
\end{equation}
from which we read off the boost and rotation Ward identities
\begin{eqnarray}
0 & = & -\hat h_{\mu\nu}T^\mu+\tau_\mu h^{\rho\sigma}\hat h_{\nu\sigma}T^{\mu}{}_{\rho}\,,\label{eq:boostWI}\\
0 & = & \hat h_{\mu\rho}\hat h_{\nu\lambda}h^{\lambda\sigma}T^{\rho}{}_\sigma-\left(\mu\leftrightarrow\nu\right)\,.\label{eq:rotationWI}
\end{eqnarray}
The definitions \eqref{eq:invtumu} and \eqref{eq:invtmunu} imply
\begin{eqnarray}
-\frac{1}{2}t^{AB}\delta \gamma_{AB} & = & -\tau_\nu T^\nu{}_\mu\delta\hat v^\mu-\left(\hat h_{\sigma\nu}\hat v^\mu T^\nu{}_\mu\right)\tau_\rho\delta h^{\rho\sigma}\nonumber\\
&&+\frac{1}{2}\left(\hat h_{\rho\nu}\hat h_{\sigma\lambda}h^{\lambda\mu}T^\nu{}_\mu\right)\delta h^{\rho\sigma}+\tau_\mu T^\mu\delta\tilde\Phi\,,
\end{eqnarray}
in agreement with the definition of $T^\mu{}_\nu$ and $T^\mu$ as the response to varying the TNC invariants $\hat v^\mu$, $h^{\mu\nu}$ and $\tilde\Phi$ as given in \eqref{eq:variationTNC}. The relation between the higher and lower dimensional energy-momentum tensors holds for any reduction given that $\gamma_{AB}$ admits a null Killing vector. No additional assumptions such as hypersurface orthogonality of $\partial_u$ are needed.

The higher-dimensional energy-momentum tensor corresponds to the boundary theory of a bulk AdS$_5$ space-time and is thus traceless. Further by boundary diffeomorphism invariance it satisfies a Ward identity for local diffeomorphism invariance. Upon reduction these give rise to Ward identities for local scale and diffeomorphism invariance. To this end it is useful to consider\footnote{Sometimes it is useful to express $t_{AB}$ in terms of lower-dimensional quantities via
\begin{eqnarray}
t_{\mu\nu}  & = & \hat h_{\rho\nu}T^\rho{}_\mu+\hat h_{\rho\mu}T^\rho{}_\nu-\hat h_{\mu\rho}\hat h_{\nu\sigma}h^{\sigma\lambda}T^\rho{}_\lambda+\tau_\mu\tau_\nu\hat v^\rho\hat v^\sigma t_{\rho\sigma}\,,\label{eq:nullredt1}\\
t_{\mu u} & = & \tau_\rho T^\rho{}_\mu\,,\label{eq:nullredt2}\\
t_{uu} & = & \tau_\rho T^\rho\,.\label{eq:nullredt3}
\end{eqnarray}
} $t^{A}{}_{B}$, i.e. 
\begin{eqnarray}
t^u{}_u & = & 2\tilde\Phi\tau_\mu T^\mu-\hat v^\nu\tau_\mu T^\mu{}_\nu\,,\\
t^u{}_\nu & = & 2\tilde\Phi\tau_\mu T^\mu{}_\nu-\hat v^\sigma\hat h_{\nu\rho}T^\rho{}_\sigma+\tau_\nu\hat v^\rho\hat v^\sigma t_{\rho\sigma}\,,\\
t^\mu{}_u & = & T^\mu\,,\label{eq:nullredt6}\\
t^\mu{}_\nu & = & T^\mu{}_\nu\,,\label{eq:nullredt7}
\end{eqnarray}
where $\hat v^\rho\hat v^\sigma t_{\rho\sigma}$, which contains $t^{uu}$, is unspecified in terms of lower dimensional quantities as it will drop out of the Ward identities. We can derive the following identities
\begin{eqnarray}
\nabla_{A} t^{A}{}_u & = & \partial_\mu\left(eT^\mu\right)\,,\\
\nabla_{A} t^{A}{}_\mu & = & e^{-1}\partial_\nu\left(e T^\nu{}_\mu\right)+T^\rho{}_\nu\left(\hat v^\nu\partial_\mu\tau_\rho-e_{a}^\nu\partial_\mu\hat e_{\rho}^{a}\right)+\tau_\nu T^\nu\partial_\mu\tilde\Phi\,,\label{eq:diffeoWI}\\
t^{A}{}_{A} & = & -2\hat v^\nu\tau_\mu T^\mu{}_{\nu}+\hat e_{\mu}^{a}e^{\nu}_a T^\mu{}_\nu+2\tilde\Phi\tau_\mu T^\mu\,.\label{eq:nullredtraceWI}
\end{eqnarray}
If we are dealing with a relativistic and scale invariant theory, i.e. $\nabla_{A} t^{A}{}_{B}=t^{A}{}_{A}=0$ we find the diffeomorphism, $U(1)$ and the $z=2$ version of the local dilatation Ward identities as given in \cite{Christensen:2013rfa,Hartong:2014oma,Hartong:2014pma,Hartong:2015wxa}.

The diffeomorphism Ward identity \eqref{eq:diffeoWI} can also be written in a TNC covariant form using the connection \eqref{eq:TNCconnection} as done in \cite{Hartong:2014oma}. Instead of the connection \eqref{eq:TNCconnection} we can also take the Riemann--Cartan connection of \cite{Hartong:2015xda}, that we will denote by $\check\Gamma^\rho_{\mu\nu}$, given by
\begin{equation}\label{eq:manifestinvGamma2TNC}
\check\Gamma^\lambda_{\mu\rho}=-\hat v^\lambda\partial_\mu\tau_\rho+\frac{1}{2}h^{\nu\lambda}\left(\partial_\mu\hat h_{\rho\nu}+\partial_\rho\hat h_{\mu\nu}-\partial_\nu\hat h_{\mu\rho}\right)-h^{\nu\lambda}\tau_\rho K_{\mu\nu}\,,
\end{equation}
where $K_{\mu\nu}=-\frac{1}{2}\mathcal{L}_{\hat v}\hat h_{\mu\nu}$ is the extrinsic curvature. This connection obeys
\begin{equation}
\check\nabla_\mu\tau_\nu=0\,,\qquad\check\nabla_\mu \hat h_{\nu\rho}=0\,,\qquad\check\nabla_\mu\hat v^\nu=0\,,\qquad\check\nabla_\mu h^{\nu\rho}=0\,,
\end{equation}
and the relation \eqref{eq:diffeoWI} becomes
\begin{equation}
\nabla_A t^A{}_\nu = \check\nabla_\mu T^\mu{}_\nu+2\check\Gamma^\rho_{[\mu\rho]}T^\mu{}_\nu-2\check\Gamma^\mu_{[\nu\rho]}T^\rho{}_\mu+\tau_\mu T^\mu\partial_\nu\tilde\Phi\,.
\end{equation}
This is the most compact and TNC covariant way of writing the diffeomorphism Ward identity. 

In \cite{Hartong:2015xda} it was shown that TTNC geometry (but not the more general TNC geometry) can be obtained by projecting the higher-dimensional metric compatibility conditions involving the Levi-Civita connection onto the surface orthogonal to $\partial_u$ (null reduction) in the sense that the TNC metric compatibility conditions follow from the projection of the higher-dimensional metric compatibility conditions only when $\partial_u$ is hypersurface orthogonal. However, the condition that the TNC metric compatibility conditions follows by projection is somewhat artificial. Here we see that the diffeomorphism Ward identity takes the required form for any field theory on a TNC geometry and not just TTNC geometry. At no point in the analysis did we assume anything about $\tau_\mu$.

\subsection{Null reduction of a perfect relativistic fluid}\label{subsec:nullredfluid}

Since we are interested in non-relativistic versions of the fluid/gravity correspondence we study here the null reduction of a relativistic fluid\footnote{This has also been done in \cite{Banerjee:2015hra} but our approach differs in that we do not need to introduce what is called a null fluid in \cite{Banerjee:2015hra}.}.

A relativistic perfect fluid is given by a conserved energy-momentum tensor $t_{AB}$ that is of the form
\begin{equation}
t_{AB}=\left(E+P\right)U_A U_B+P\gamma_{AB}\,,
\end{equation}
where $U_A$ satisfies $U_A U^A=-1$. Consider the following parametrization of $U_A$, 
\begin{eqnarray}
U_u^2 & = & \frac{\rho}{E+P}\,,\\
h^{\mu\nu}U_\nu & = & U_u\left(\hat v^\mu-u^\mu\right)\,,\\
\hat v^\mu U_\mu & = & \frac{1}{2}U_u\left(\hat h_{\mu\nu}u^\mu u^\nu+2\tilde\Phi+U_u^{-2}\right)\,,
\end{eqnarray}
where $u^\mu$ satisfies $\tau_\mu u^\mu=-1$. It follows that
\begin{eqnarray}
U_\mu & = & -\frac{1}{2}U_u\tau_\mu\left(\hat h_{\rho\sigma}u^\rho u^\sigma+2\tilde\Phi+U_u^{-2}\right)-U_u\hat h_{\mu\nu}u^\nu\\
& = & -U_u\left[\frac{1}{2}\tau_\mu\left(h_{\rho\sigma} u^\rho u^\sigma+U_u^{-2}\right)+h_{\mu\nu}u^\nu+m_\mu\right]\,,
\end{eqnarray}
which (except for the $U_u^{-2}$ term) takes the form of the velocity of a point particle. The components of $U^\mu$ are given by
\begin{eqnarray}
U^u & = & -\frac{1}{2}U_u\left(\hat h_{\mu\nu}u^\mu u^\nu-2\tilde\Phi+U_u^{-2}\right)\,,\\
U^\mu & = & -U_u u^\mu\,.\label{eq:fluidvelo}
\end{eqnarray}
Further redefine the energy density $E$ as
\begin{equation}
E=2\mathcal{E}+P\,.
\end{equation}
Using the above results it follows that $T^\mu{}_\nu$ and $T^\mu$ are given by
\begin{eqnarray}
T^\mu{}_\nu & = & \left(\mathcal{E}+P+\rho\tilde\Phi+\frac{1}{2}\rho\hat h_{\lambda\kappa}u^\lambda u^\kappa\right)u^\mu\tau_\nu+P\delta^\mu_\nu+\rho u^\mu\hat h_{\nu\rho}u^\rho\nonumber\\
& = & \left(\mathcal{E}+P+\frac{1}{2}\rho h_{\lambda\kappa}u^\lambda u^\kappa\right)u^\mu\tau_\nu+P\delta^\mu_\nu+\rho u^\mu h_{\nu\rho}u^\rho+\rho u^\mu m_\nu\,,\label{eq:PFEMT3}\\
T^\mu & = & -\rho u^\mu\,.\label{eq:PFEMT4}
\end{eqnarray}
This can be shown to agree with the notion of a Galilean perfect fluid as given in \cite{Jensen:2014ama}.

The null reduction ansatz has a local $U(1)$ symmetry which is the diffeomorphism $\delta u=-\xi^u$ and $\delta m_\mu=-\partial_\mu\xi^u$. If we act with this diffeomorphism on $U^A$ and $U_A$ via
\begin{equation}
\delta U_A=\xi^B\partial_B U_A+U_B\partial_A\xi^B\,,\qquad\delta U^A=\xi^B\partial_B U^A-U^B\partial_B\xi^A\,,
\end{equation}
with $\xi^A=\delta^A_u\xi^u$ we see that $U_u$ and $U^\mu$ are $U(1)$ invariant. It follows from \eqref{eq:fluidvelo} that the fluid velocity $u^\mu$ is particle number invariant. 

The $z=2$ trace Ward identity reads (for $d=2$ spatial dimensions)
\begin{equation}
 t^{A}{}_{A} = -2\hat v^\nu\tau_\mu T^\mu{}_{\nu}+\hat e_{\mu}^{a}e^{\nu}_a T^\mu{}_\nu+2\tilde\Phi\tau_\mu T^\mu=-2\mathcal{E}+2P=0\,.
\end{equation}
The null reduction only leads to theories with $z=2$ scaling relations.

\section{Holographic renormalization of the upliftable model}\label{app:dimred}

As discussed in section \ref{subsec:dimred} the EPD model with
\begin{equation}
Z=e^{3\Phi}\,,\qquad W=4\,,\qquad V(\Phi)=2e^{-3\Phi}-12e^{-\Phi}\,,\qquad x=3\,.
\end{equation}
can be obtained from a Scherk--Schwarz reduction of the 5-dimensional action
\begin{equation}\label{eq:uplifted2}
S=\frac{1}{2\kappa_5^2}\int d^5x\sqrt{-\mathcal{G}}\left(R+12-\frac{1}{2}\partial_{\mathcal{M}}\psi\partial^{\mathcal{M}}\psi\right)\,,
\end{equation}
where $\kappa_5^2=8\pi G_5$ with $G_5$ the 5-dimensional Newton's constant and where $\mathcal{M}=(u,M)$. The consistency of this reduction will be shown in section \ref{subsec:consistency}. 

In this appendix we will first perform the holographic renormalization in 5 dimensions for those asymptotically locally AdS space-times that have a boundary metric obeying the null reduction ansatz of section \ref{subsec:nullred}. We then subsequently reduce the result to obtain the counterterms and near-boundary expansions in 4 dimensions for asymptotically locally $z=2$ Lifshitz space-times.

\subsection{Fefferman--Graham expansions and counterterms}\label{sec:FGexpansions}

By using the results of \cite{deHaro:2000xn,Papadimitriou:2011qb,Chemissany:2012du,Christensen:2013rfa}{}\footnote{We set $\hat\chi=0$ and redefine $\hat\phi=\psi$ in \cite{Christensen:2013rfa}.} we can obtain the solution to the equations of motion of \eqref{eq:uplifted2} (that are given further below in \eqref{eq:Einsteineqs} and \eqref{eq:phieom}) expressed as an asymptotic series in radial gauge, i.e. as a Fefferman--Graham (FG) expansion \cite{FeffermanGraham}. The result reads\footnote{We will denote here and further below by $a_{(n,m)}$ the coefficient at order $r^n(\log r)^m$ of the field $r^\Delta a$ where $r^{-\Delta}$ is the leading term in the expansion of $a$ with the exception of the $a_{(n,0)}$ term which we will simply denote as $a_{(n)}$.}
\begin{eqnarray}
\mathcal{G}_{\mathcal{M}\mathcal{N}}dx^{\mathcal{M}} dx^{\mathcal{N}} &=& \frac{dr^2}{r^2}+\gamma_{AB}dx^{A}dx^{B}\,, \label{eq: sol metric gauge} \\
\gamma_{AB} &=& \frac{1}{r^2}\left[\gamma_{(0)AB}+r^2\gamma_{(2)AB}+r^4\log r \gamma_{(4,1)AB}+r^4\gamma_{(4)AB}+O(r^6\log r)\right]\,, \label{eq: sol metric}\\
\psi &=& \psi_{(0)} + r^2\psi_{(2)}+r^4\log r\psi_{(4,1)} + r^4\psi_{(4)}+O(r^6\log r)\,, \label{eq: sol phi}
\end{eqnarray}
where the coefficients are given by
\begin{eqnarray}
\gamma_{(2)AB} &=& -\frac{1}{2}\left(R_{(0)AB}-\frac{1}{2}\partial_{A}\psi_{(0)}\partial_{B}\psi_{(0)}\right)+\frac{1}{12}\gamma_{(0)AB}\left(R_{(0)}-\frac{1}{2}(\partial\psi_{(0)})^2\right)\,,\label{eq:h2ab}\\
\psi_{(2)} &=& \frac{1}{4}\square_{(0)}\psi_{(0)} \,,
\end{eqnarray}
at second order and by
\begin{eqnarray}
\gamma_{(4,1)AB} &=&  \frac{1}{4}\nabla^{C}_{(0)}\left(\nabla_{(0)A}\gamma_{(2)BC}+\nabla_{(0)B}\gamma_{(2)AC}-\nabla_{(0)C}\gamma_{(2)AB}\right) -\frac{1}{4}\nabla_{(0)A}\nabla_{(0)B}\gamma^{C}_{(2)C} \nonumber\\
&& +\gamma_{(2)AC}\gamma^{C}_{(2)B}-\frac{1}{2}\partial_{(A}\psi_{(0)}\nabla_{(0)B)}\psi_{(2)}    -\gamma_{(0)AB}\left( \frac{1}{4}\gamma_{(2)}^{CD}\gamma_{(2)CD}+\frac{1}{2}\psi_{(2)}^2\right) \,,\\
\psi_{(4,1)} &=& -\frac{1}{4}\left[\square_{(0)}\psi_{(2)}+2\psi_{(2)}\gamma^{A}_{(2)A} +\frac{1}{2}\partial^{A}\psi_{(0)}\nabla_{(0)A}\gamma^{B}_{(2)B}-\gamma^{AB}_{(2)}\nabla_{(0)A}\partial_{B}\psi_{(0)} \right.\nonumber\\
&& \left. -\partial^{A}\psi_{(0)}\nabla^{B}_{(0)}\gamma_{(2)AB}\right]\,,
\end{eqnarray}
at order $r^4\log r$. We note that the quantity $\gamma_{(4,1)AB}$ is traceless.  Indices of the expansion coefficients are raised and lowered with the AdS boundary metric $\gamma_{(0)AB}$. At order $r^4$ we have that $\gamma_{(4)AB}$ is constrained by
\begin{eqnarray}
\gamma_{(4)A}^{A} & = & \frac{1}{4}\gamma_{(2)AB}\gamma_{(2)}^{AB}-\frac{1}{2}\psi_{(2)}^2\,,\label{eq:traceh4}\\
\nabla^{B}_{(0)}\gamma_{(4)AB} & = & \psi_{(4)}\partial_{A}\psi_{(0)}-\frac{1}{2}\psi_{(2)}\nabla_{(0)A}\psi_{(2)}-\frac{1}{4}\gamma_{(2)}^{BC}\nabla_{(0)A}\gamma_{(2)BC}\nonumber\\
&&-\frac{1}{4}\gamma_{(2)AC}\nabla^{C}_{(0)}\gamma_{(2)B}^{B}+\frac{1}{2}\gamma_{(2)}^{BC}\nabla_{(0)B}\gamma_{(2)AC}+\frac{1}{2}\gamma_{(2)A}^{C}\nabla^{B}_{(0)}\gamma_{(2)BC}\,.\label{eq:divh4}
\end{eqnarray}
Following \cite{deHaro:2000xn} we write the coefficient $\gamma_{(4)AB}$ as
\begin{equation}
\gamma_{(4)AB} =X_{AB} - \frac{1}{4}t_{AB}\,,\label{eq: h4ab}
\end{equation}
where $t_{AB}$ is the boundary energy-momentum tensor defined in \eqref{eq:bdrystresstensor}.
The trace and divergence of $t_{AB}$ will be given below together with the explicit form of $X_{AB}$. In the expansion for the scalar we have that $\psi_{(4)}$ is a fully arbitrary function of the boundary coordinates.

The complete action with Gibbons--Hawking and local counterterms (using minimal subtraction) is given by
\begin{equation}\label{eq:actionmodel}
S_{\text{ren}} =\frac{1}{2\kappa_5^2}\int_{\mathcal{M}}d^5x\sqrt{-\mathcal{G}}\left(R+12-\frac{1}{2}\partial_{\mathcal{M}}\psi\partial^{\mathcal{M}}\psi\right)
+\frac{1}{\kappa_5^2}\int_{\partial\mathcal{M}}d^4x\sqrt{-\gamma}K+S_{\text{ct}}\,,
\end{equation}
where $\gamma$ denotes the determinant of the metric $\gamma_{AB}$ on the cut off boundary $\partial\mathcal{M}$, the extrinsic curvature $K$ is given by
\begin{equation}
K=\gamma^{AB}K_{AB}\,,\qquad K_{AB}=-\frac{1}{2}\mathcal{L}_n \gamma_{AB}\,,\qquad n^{\mathcal{M}}=-r\delta^{\mathcal{M}}_r\,,
\end{equation}
and where
\begin{equation}\label{eq:Sct1}
S_{\text{ct}}=\frac{1}{\kappa_5^2}\int_{\partial\mathcal{M}}d^4x\sqrt{-\gamma}\left(-\frac{1}{4}\left(R_{(\gamma)} +12 - \frac{1}{2}\partial_{A}\psi\partial^A\psi\right)-\frac{1}{2}\mathcal{A}\log r\right)\,,
\end{equation}
with
\begin{eqnarray}
\mathcal{A} &=& -\frac{1}{4}\left(Q^{AB}Q_{AB}-\frac{1}{3}Q^2+\frac{1}{2}\left(\square_{(\gamma)}\psi\right)^2\right)\,,\label{eq:anomalycounterterm}\\
Q_{AB} & = & R_{(\gamma)AB} - \frac{1}{2}\partial_{A}\psi\partial_{B}\psi\,.\nonumber
\end{eqnarray}

\subsection{One-point functions}\label{subsec:onepointfunctions}

To compute one-point functions, we write the total variation of $S_{\text{ren}}=S_{\text{bulk}}+S_{\text{GH}}+S_{\text{ct}}$ as
\begin{eqnarray}
\delta S_{\text{ren}} &=&
\frac{1}{2\kappa_5^2}\int_{\mathcal{M}}d^5x\sqrt{-\mathcal{G}}\left(
\mathcal{E}_{\mathcal{M}\mathcal{N}}\delta \mathcal{G}^{\mathcal{M}\mathcal{N}} + \mathcal{E}_{\psi}\delta\psi
 \right) \nonumber\\ && +
\frac{1}{2\kappa_5^2} \int_{\partial\mathcal{M}}d^4x\sqrt{-\gamma} \left(
\frac{1}{2}T_{AB}\delta\gamma^{AB} + T_{\psi}\delta\psi  \right)\,,\label{eq:deltaS}
\end{eqnarray}
where $\mathcal{E}_{\mathcal{M}\mathcal{N}}$ and $\mathcal{E}_{\psi}$
are the equations of motion 
\begin{eqnarray}
\mathcal{E}_{\mathcal{M}\mathcal{N}} & = & G_{\mathcal{M}\mathcal{N}}-6\mathcal{G}_{\mathcal{M}\mathcal{N}}-\frac{1}{2}\partial_{\mathcal{M}}\psi\partial_{\mathcal{N}}\psi+\frac{1}{4}\mathcal{G}_{\mathcal{M}\mathcal{N}}(\partial\psi)^2\,,\label{eq:Einsteineqs}\\
\mathcal{E}_{\psi} & = & \square\psi\,,\label{eq:phieom}
\end{eqnarray}
and where
\begin{eqnarray}
T_{AB} &=& -2(K-3)\gamma_{AB} +2K_{AB} 
-Q_{AB}+ \frac{1}{2}h_{AB}Q + \log r T^{(A)}_{AB} \,,\label{eq:Tab}\\
T_{\psi} &=& -n^{M}\partial_{M}\psi -
\frac{1}{2}\square_{(\gamma)}\psi +\log r T^{(A)}_{\psi} \label{eq:Tphi}\,.
\end{eqnarray}
Here we defined
\begin{equation}
T^{(A)}_{AB} = - \frac{2\kappa_5^2}{\sqrt{-\gamma}} \frac{\delta A}{\delta
\gamma^{AB}}\,, \qquad T^{(A)}_{\psi} = - \frac{\kappa_5^2}{\sqrt{-\gamma}}
\frac{\delta A}{\delta\psi}\,,
\end{equation}
with
\begin{equation}
A=\frac{1}{\kappa_5^2}\int_{\partial\mathcal{M}}d^4x\sqrt{-\gamma}\mathcal{A}\,.
\end{equation}

From the expansions it follows that
$\sqrt{-\gamma}=r^{-4}\sqrt{-\gamma_{(0)}}+O(r^{-2})$, $\delta
\gamma^{AB} = r^2\delta\gamma_{(0)}^{AB} + O(r^4)$, $\delta\psi =
\delta\psi_{(0)} + O(r^2)$,
which is used to  obtain the following one-point functions (we take the cut-off boundary at $r=\epsilon$)
\begin{eqnarray}
t_{AB} &=&  \frac{4\kappa_5^2}{\sqrt{-\gamma_{(0)}}} \frac{\delta S_{\text{ren}}^{\text{on-shell}}}{\delta\gamma_{(0)}^{AB}} = \lim_{\epsilon \rightarrow 0} \epsilon^{-2}T_{AB} = -4\gamma_{(4)AB} +4 X_{AB}\,,\label{eq:bdrystresstensor}\\
\langle\mathcal{O}_{\psi} \rangle &=&  \frac{2\kappa_5^2}{\sqrt{-\gamma_{(0)}}} \frac{\delta S_{\text{ren}}^{\text{on-shell}}}{\delta\psi_{(0)}} = \lim_{\epsilon \rightarrow 0} \epsilon^{-4}T_{\psi} = 4\psi_{(4)}+\psi_{(2)}\gamma^{A}_{(2)A} +3\psi_{(4,1)}\,,
\end{eqnarray}
where
\begin{equation}
X_{AB} = \frac{1}{2}\gamma_{(2)AC}\gamma^{C}_{(2)B}-\frac{1}{4}\gamma^{C}_{(2)C}\gamma_{(2)AB} +
\frac{1}{8}\gamma_{(0)AB}\mathcal{A}_{(0)} - \frac{3}{4}\gamma_{(4,1)AB} \,,\label{eq: Xab}
\end{equation}
with
\begin{equation}\label{eq:hatA0}
\mathcal{A}_{(0)}=\lim_{\epsilon \rightarrow 0}
\epsilon^{-4}\mathcal{A}
=(\gamma_{(2)A}^{A})^2-\gamma_{(2)}^{AB}\gamma_{(2)AB}-2\psi_{(2)}^2\,.
\end{equation}

Using equations \eqref{eq:traceh4} and \eqref{eq:divh4} we can compute
the trace and divergence of the boundary energy-momentum tensor and the result is
\begin{eqnarray}
t^{A}{}_{A} &=& \mathcal{A}_{(0)}\,,\label{eq:tracet}\\
\nabla_{(0)A}t^{A}{}_{B} &=& -\langle \mathcal{O}_{\psi} \rangle\partial_{B}\psi_{(0)} \,.\label{eq:divt}
\end{eqnarray}

\subsection{Dimensional Reduction of the action}\label{subsec:dimredTNC}

The Scherk--Schwarz reduction leading to \eqref{eq:action} with the choices \eqref{eq:upliftablemodel} is obtained by the following reduction ansatz
\begin{eqnarray}
	ds_5^{2} &=& \mathcal{G}_{\mathcal{M}\mathcal{N}} dx^{\mathcal{M}}dx^{\mathcal{N}} = \frac{dr^{2}}{r^2} + \gamma_{AB}dx^{A}dx^{B} = e^{-\Phi}g_{MN}dx^{M}dx^{N} + e^{2\Phi}\left(du + A_{M}dx^{M}  \right)^2 \nonumber \\
	& =&e^{-\Phi} \left( e^{\Phi}\frac{dr^{2}}{r^2} + h_{\mu\nu}dx^{\mu}dx^{\nu}\right) + e^{2\Phi}\left(du + A_{\mu}dx^{\mu} \right)^2\,,\label{eq:KK} \\
	\psi &=& 2u+2\Xi\,, \label{eq:KK2}
\end{eqnarray}
where all the functions are independent of the fifth coordinate $u$ which is periodically identified, so $u \sim u + 2\pi L$. The only exception is the term linear in $\psi$ which means that upon going around the reduction circle $\psi$ comes back to itself up to a constant shift. This is allowed because shifting $\psi$ is a global symmetry of the higher-dimensional theory. The consistency of the reduction is proven in section \ref{subsec:consistency}.

After reduction the four dimensional action is 
\begin{eqnarray}
	S & = &  \int d^{4}x \sqrt{-g}\left(R - \frac{3}{2}\partial_M\Phi\partial^M\Phi - \frac{1}{4}e^{3\Phi}F_{MN}F^{MN} - 2B_M B^M - V \right) \nonumber \\ 
	&& +2 \int d^{3}x\sqrt{-h} K + S_{\text{ct}}\,, \label{eq:4Daction}\\
	S_{\text{ct}} & = & 2\int_{\partial \mathcal{M}} d^3 x\sqrt{-h}\left[ - \frac{1}{4}e^{\Phi/2}\left(R_{(h)}
	 - \frac{3}{2}\partial_\mu \Phi\partial^\mu\Phi - \frac{1}{4}e^{3\Phi}F_{\mu\nu}F^{\mu\nu}   - 2B_\mu B^\mu +10e^{-\Phi} \right) \right] \nonumber \\
	& &- \log r\int_{\partial \mathcal{M}} d^3 x\sqrt{-h}e^{-\Phi /2}\mathcal{A}\,,\label{eq:counterterms4D}
\end{eqnarray}
where
\begin{eqnarray}
	B_{M} & = &  A_{M}-\partial_{M}\Xi\,, \\
	F_{MN} & = & \partial_{M}B_{N} - \partial_{N}B_{M}\,, \\
	V & = & 2e^{-3\Phi} - 12e^{-\Phi}\,,
\end{eqnarray}
and where we used that $\frac{2\pi L}{2\kappa_{5}^{2}}=1$.

The total variation can be written as
\begin{eqnarray}
 \delta S_{ren} & = & \int_{\mathcal{M}}d^{4}x \sqrt{-g}\left( \mathcal{E}_{MN} \delta g^{MN} + \mathcal{E}^{N}\delta B_{N}+ \mathcal{E}_{\Phi}\delta \Phi  \right) \nonumber \\ 
 && +\int_{\partial \mathcal{M}}d^{3}x\sqrt{-h}\left(\frac{1}{2}T_{\mu\nu}\delta h^{\mu\nu}  + \mathcal{T}^{\nu} \delta B_{\nu}+ T_{\Phi} \delta \Phi \right)\,, \label{totalvariation}
\end{eqnarray}
with
\begin{eqnarray}
 \mathcal{E}_{MN} & = &G_{MN}  + \frac{1}{8}e^{3\Phi}g_{MN}F_{PQ}F^{PQ}  - \frac{1}{2}e^{3\Phi}F_{MP}F_{N}{}^{P}  + g_{MN}B_P B^P - 2B_{M}B_{N}\nonumber \\ 
 & &+ \frac{3}{4}g_{MN}\partial_P\Phi\partial^P\Phi- \frac{3}{2}\partial_{M}\Phi \partial_{N}\Phi  + \frac{1}{2}g_{MN} V\,, \label{eq:eomupliftable1}\\
 \mathcal{E}_{\Phi} & = & 3\square \Phi - \frac{3}{4}e^{3\Phi}F_{MN}F^{MN} + 6e^{-3\Phi } - 12e^{-\Phi}\,, \label{eq:eomupliftable2}\\
 \mathcal{E}^{N} & = & \nabla_{M}\left(e^{3\Phi}F^{MN} \right) - 4B^{N}\,,\label{eq:eomupliftable3}
\end{eqnarray}
and
\begin{eqnarray}
 T_{\mu\nu} & = & -2Kh_{\mu\nu} +2 K_{\mu\nu}-e^{\Phi/2}G_{(h)\mu\nu}+5e^{ - \Phi/2}h_{\mu\nu}\nonumber\\
 &&+ \frac{1}{2}e^{7\Phi/2} F_{\mu\rho}F_{\nu}{}^{\rho}- \frac{1}{8}e^{7\Phi/2}h_{\mu\nu}F_{\rho\sigma}F^{\rho\sigma} -e^{\Phi/2}h_{\mu\nu}B_\rho B^\rho+2e^{\Phi/2}B_{\mu}B_{\nu} \nonumber \\ 
 &&+ \frac{1}{2}e^{\Phi/2}\left( \nabla^{(h)}_\mu\partial_{\nu} \Phi - h_{\mu\nu}\square_{(h)} \Phi \right)  + \frac{7}{4}e^{\Phi/2} \partial_{\mu}\Phi \partial_{\nu} \Phi -e^{\Phi/2}h_{\mu\nu}\partial_\rho\Phi\partial^\rho\Phi\,, \label{eq:Tmunu}\\ 
 T_{\Phi} & = & -3n^{M}\partial_{M} \Phi - \frac{1}{4}e^{\Phi/2}R_{(h)}  - \frac{3}{8}e^{\Phi/2}\partial_\mu\Phi\partial^\mu\Phi - \frac{3}{2}e^{\Phi/2}\square_{(h)} \Phi \nonumber \\ 
 &&  + \frac{7}{16}e^{7\Phi/2}F_{\mu\nu}F^{\mu\nu}+ \frac{1}{2}e^{\Phi/2}B_\mu B^\mu + \frac{5}{2}e^{-\Phi/2}\,, \\
  \mathcal{T}^{\nu} & = & -e^{3\Phi}n_{M}F^{M\nu} - \frac{1}{2}\nabla^{(h)}_{\mu}\left(e^{7\Phi/2}F^{\mu\nu}\right) +2 e^{ \Phi/2} B^{\nu}\,,\label{eq:calT}
\end{eqnarray}
where the extrinsic curvature is 
\begin{equation}
K=h^{\mu\nu}K_{\mu\nu}\,,\qquad K_{\mu\nu}=-\frac{1}{2}\mathcal{L}_n h_{\mu\nu}\,,\qquad n^M=-re^{-\Phi/2}\delta^M_r\,.
\end{equation}
These expressions are correct up to $\log r$ terms since we did not vary those counterterms.

\subsection{Sources and Vevs}\label{subsec:sourcesvevs}

We write the 4-dimensional metric in \eqref{eq:KK} as
\begin{equation}\label{eq:vielbeindecomp}
ds^2 =e^{\Phi}\frac{dr^{2}}{r^2} + h_{\mu\nu}dx^{\mu}dx^{\nu}=e^{\Phi} \frac{dr^2}{r^2}-E^0 E^0+\delta_{ab}E^a E^b\,.
\end{equation}
In order to compute the vevs we use the identity \cite{Christensen:2013rfa,Hartong:2014oma}
\begin{equation}\label{eq:variation}
\frac{1}{2}T_{\mu\nu}\delta h^{\mu\nu}  + \mathcal{T}^{\nu} \delta B_{\nu}+ T_{\Phi} \delta \Phi=\mathcal{S}^0_\mu\delta E^\mu_0+\mathcal{S}^a_\mu\delta E^\mu_a+\mathcal{T}_\varphi\delta\varphi+\mathcal{T}^a\delta A_a+\mathcal{T}_\Xi\delta\Xi+\mathcal{T}_\Phi\delta\Phi\,,
\end{equation}
which holds up to a total derivative, where we used that $B_\nu=A_\nu-\partial_\nu\Xi$, $A_a=E^\mu_a A_\mu$ and where $\varphi$ is defined by \cite{Hartong:2014oma}
\begin{equation}
\varphi = E_0^\nu A_\nu-\alpha(\Phi)\,,
\end{equation}
with $\alpha=e^{-3\Phi/2}$ for the particular model studied here \cite{Christensen:2013rfa} and where 
\begin{eqnarray}
\mathcal{S}^0_\mu & = & -\left(T_{\mu\nu}E^\nu_0+\mathcal{T}^\rho E_\rho^0 A_\mu\right)\,,\label{eq:S0}\\
\mathcal{S}^a_\mu & = & \left(T_{\mu\nu}E^{\nu a}-\mathcal{T}^\rho E_\rho^a A_\mu\right)\,,\label{eq:Sa}\\
\mathcal{T}_\varphi & = & \mathcal{T}^\nu E_\nu^0\,,\\
\mathcal{T}_\Phi & = & T_\Phi+\mathcal{T}^\nu E_\nu^0\frac{d\alpha}{d\Phi}\,,\\
\mathcal{T}^a & = & \mathcal{T}^\nu E_\nu^a\,,\\
\mathcal{T}_\Xi & = & e^{-1}\partial_\mu\left(e\mathcal{T}^\mu\right)\,.
\end{eqnarray}

The 4-dimensional sources are defined as the leading terms in the expansions of the bulk fields appearing on the right hand side of \eqref{eq:variation}. We find the sources $v^\mu, e^\mu_a, m_\mu, \phi, \chi$ defined via
\begin{eqnarray}
E_0^\mu & \simeq & -r^2\alpha_{(0)}^{-1/3}v^\mu\,,\label{eq:sources1}\\
E_a^\mu & \simeq & r\alpha_{(0)}^{1/3}e^\mu_a\,,\label{eq:sources2}\\
A_\mu-\alpha(\Phi)E^0_\mu & \simeq & -m_\mu\,,\label{eq:sources3}\\
\Phi & \simeq & \phi\,,\label{eq:sources4}\\
\Xi & \simeq & -\chi\,,\\
\varphi & \simeq & r^2\alpha_{(0)}^{-1/3} v^\mu m_\mu\,,\\
A_a & \simeq & -r\alpha_{(0)}^{1/3} e^\mu_a m_\mu\,.
\end{eqnarray}
Likewise the vevs are defined as the leading terms in the expansions of the objects that are the responses to the variations written in \eqref{eq:variation}, i.e. we define the vevs $S^0_\mu, S^a_\mu, T^0, T^a, \langle O_\phi\rangle, \langle O_\chi\rangle$
\begin{eqnarray}
\mathcal{S}^0_\mu & \simeq & r^2\alpha_{(0)}^{2/3}S^0_\mu\,,\label{eq:vev1}\\
\mathcal{S}^a_\mu & \simeq & r^3S^a_\mu\,,\label{eq:vev2}\\
\mathcal{T}_\varphi & \simeq & -r^2\alpha_{(0)}^{2/3}T^0\,,\\
\mathcal{T}^a & \simeq & -r^3T^a\,,\\
\mathcal{T}_\Phi & \simeq & r^4\alpha_{(0)}^{1/3}\langle O_\phi\rangle\,,\\
\mathcal{T}_\Xi & \simeq & -r^4\alpha_{(0)}^{1/3}\langle O_\chi\rangle\,,
\end{eqnarray}
where 
\begin{equation}\label{eq:alpha0}
\alpha_{(0)}=e^{-3\phi/2}\,.
\end{equation}

Using \eqref{eq:vielbeindecomp}, \eqref{eq:KK} as well as the definitions of the 4-dimensional sources \eqref{eq:sources1}--\eqref{eq:sources3} we can derive the following relation between the 5-dimensional boundary metric $\gamma_{(0)AB}$ and the 4-dimensional sources $\tau_\mu$, $m_\mu$ and $e_\mu^a$,
\begin{equation}\label{eq:nullredhigherDmetric}
ds^2=\gamma_{(0)AB}dx^{A}dx^{B}=2\tau_\mu dx^\mu\left(du-m_\nu dx^\nu\right)+h_{\mu\nu}dx^\mu dx^\nu\,,
\end{equation}
where $h_{\mu\nu}=\delta_{ab}e^a_\mu e^b_\nu$ which is the form of a null reduction ansatz for a reduction along $u$ as discussed in section \ref{subsec:nullred}. The fact that the boundary metric of the 5-dimensional asymptotically locally AdS space-time must have a null circle means that the source $\phi$ which appears in the expansion of $\Phi$ is not independent of the other sources. This can be seen by noting that $\gamma_{uu}=e^{2\Phi}$, so that the 5-dimensional FG expansion via \eqref{eq: sol metric} and \eqref{eq:h2ab} tells us that 
\begin{equation}\label{eq:constraint}
e^{2\phi}=\gamma_{(2)uu}=-\frac{1}{2}R_{(0)uu}+1=-\frac{1}{4}\left(\epsilon^{\mu\nu\rho}\tau_\mu\partial_\nu\tau_\rho\right)^2+1\,,
\end{equation}
where the epsilon tensor is given by $\epsilon^{\mu\nu\rho}=e^{-1} \varepsilon^{\mu\nu\rho}$ where $e$ is the determinant of the TNC vielbein matrix $(\tau_\mu\,, e^a_\mu)$ and $\varepsilon^{\mu\nu\rho}$ is the Levi-Civita symbol. For more details we refer to \cite{Christensen:2013rfa}. The consequence of this is that the variation of the on-shell with respect to $\phi$ gives zero since nothing depends on $\phi$.

We now relate the 5-dimensional vevs to the 4-dimensional vevs. For all solutions obeying the reduction ansatz the variation of the on-shell action can be written in both a 5-dimensional and a 4-dimensional notation. From a 5-dimensional perspective we have 
\begin{eqnarray}
\delta S_{\text{ren}}^{\text{on-shell}} & = & \lim_{\epsilon\rightarrow 0}
\frac{1}{2\kappa_5^2} \int_{r=\epsilon}d^4x\sqrt{-\gamma} \left(
\frac{1}{2}T_{AB}\delta\gamma^{AB} + T_{\psi}\delta\psi  \right)\nonumber\\
&=&\int_{\partial\mathcal{M}}d^3 x e\left(\frac{1}{2}t_{AB}\delta\gamma_{(0)}^{AB}+\langle O_\psi\rangle\delta\psi_{(0)}\right)\,,\label{eq:var1}
\end{eqnarray}
where we used the fact that $\sqrt{-\gamma_{(0)}}=e=\text{det}\left(\tau_\mu,e_\mu^a\right)$ as follows from \eqref{eq:nullredhigherDmetric} and the fact that nothing depends on $u$ so that we can perform the $u$ integral. At the same time from a 4-dimensional perspective we also have, using \eqref{totalvariation}, \eqref{eq:variation},
\begin{eqnarray}
\delta S_{\text{ren}}^{\text{on-shell}} & = & \lim_{\epsilon\rightarrow 0} \int_{r=\epsilon}d^{3}x\sqrt{-h}\left(\frac{1}{2}T_{\mu\nu}\delta h^{\mu\nu}  + \mathcal{T}^{\nu} \delta B_{\nu}+ T_{\Phi} \delta \Phi \right)\label{eq:5Dosvar}\\
&=&\int_{\partial\mathcal{M}}d^3 x e\left(-S_\mu^0\delta v^\mu+S_\mu^a\delta e^\mu_a+T^0\delta m_0+T^a\delta m_a+\langle O_\chi\rangle\delta\chi+\langle\tilde O_\phi\rangle\delta\phi\right)\nonumber\,,
\end{eqnarray}
with $m_0=-v^\mu m_\mu$, $m_a=e^\mu_a m_\mu$ and where we used 
\begin{equation}\label{eq:5d4dsource}
\psi_{(0)}=2u-2\chi\,,\qquad \langle O_\psi\rangle =-\tfrac{1}{2}\langle O_\chi\rangle\,,
\end{equation}
so that $\delta\psi_{(0)}=-2\delta\chi$ and where furthermore $\tilde O_\phi$ is given by
\begin{equation}
\tilde O_\phi=O_\phi-\frac{1}{2}\left[v^\mu\left(S_\mu^0+T^0 m_\mu\right)+e^\mu_a\left(S^a_\mu+T^a m_\mu\right)\right]=0\,,
\end{equation}
which must vanish because of the comment below \eqref{eq:constraint}. The extra terms added to $O_\phi$ come from the variation of $\phi$ due to the $\alpha_{(0)}(\phi)$ factors in \eqref{eq:sources1}--\eqref{eq:sources3}. Equating \eqref{eq:5Dosvar} with \eqref{eq:var1} we obtain
\begin{equation}
\frac{1}{2}t_{AB}\delta\gamma_{(0)}^{AB}+\langle O_\psi\rangle\delta\psi_{(0)}=-S_\mu^0\delta v^\mu+S_\mu^a\delta e^\mu_a+T^0\delta m_0+T^a\delta m_a+\langle O_\chi\rangle\delta\chi+\langle\tilde O_\phi\rangle\delta\phi\,,
\end{equation}
up to total derivatives. The right hand side can be rewritten as follows
\begin{align}
&-S_\mu^0\delta v^\mu+S_\mu^a\delta e^\mu_a+T^0\delta m_0+T^a\delta m_a+\langle O_\chi\rangle\delta\chi+\langle\tilde O_\phi\rangle\delta\phi =\\
&-\tau_\nu T_\chi{}^\nu{}_\mu\delta\hat v_\chi^\mu-\left(\tau_{(\mu}\hat h^\chi_{\nu)\rho}\hat v_\chi^\sigma T_\chi{}^\rho{}_\sigma\right)\delta h^{\mu\nu}+\frac{1}{2}\left(\hat h^\chi_{\mu\rho}\hat h^\chi_{\nu\lambda}h^{\lambda\sigma} T_\chi{}^\rho{}_\sigma\right)\delta h^{\mu\nu}+\tau_\mu T^\mu\delta\tilde\Phi_\chi\nonumber\\
&+\left(\langle O_\chi\rangle-\frac{1}{e}\partial_\mu\left(eT^\mu\right)\right)\delta\chi+\left(\hat e_\chi{}_\mu^a T^\mu-\tau_\nu e^{\mu a}T_\chi{}^\nu{}_\mu\right)\delta M_a-\frac{1}{2}\hat e_\chi{}^{[a}_\nu e^{b]\mu}T_\chi{}^\nu{}_\mu\left(\hat e_\chi{}_{\rho a}\delta e^\rho_b-\hat e_\chi{}_{\rho b}\delta e^\rho_a\right)\,,\nonumber
\end{align}
where $\hat v_\chi^\mu$, $\hat e_\chi{}_\mu^a$ and $\tilde\Phi_\chi$ are given by \eqref{eq:tildePhi}, \eqref{eq:hatv} and \eqref{eq:hate} but with $m_\mu$ replaced by $M_\mu$ which is 
\begin{equation}
M_\mu=m_\mu-\partial_\mu\chi\,.
\end{equation}
This does not affect their orthonormality properties. Further we defined $M_a=e^\mu_a M_\mu$ and
\begin{eqnarray}
T_\chi{}^\mu{}_\nu & = & -\left(S^0_\nu+T^0\partial_\nu\chi\right)v^\mu+\left(S^a_\nu+T^a\partial_\nu\chi\right)e^\mu_a\,,\label{eq:chiT}\\
T^\mu & = & -T^0 v^\mu+T^a e_a^\mu\,.
\end{eqnarray}
The definitions of the 4-dimensional sources \eqref{eq:nullredhigherDmetric} and \eqref{eq:5d4dsource} imply that they transform under the local symmetries 
as TNC fields
\begin{eqnarray}
\delta e_\mu^a & = & \tau_\mu\lambda^a + \lambda^{a}{}_b e_{\mu}^{b}  \,,\\
\delta m_\mu & = & \partial_\mu\sigma+\lambda_a e^a_\mu\,,\\
\delta\chi & = & \sigma\,,
\end{eqnarray}
for the same reasons as discussed in section \ref{subsec:nullred}. From this we conclude that 
\begin{eqnarray}
\delta e^\mu_a & = & \lambda_a{}^b e^\mu_b  \,,\\
\delta M_a & = & \lambda_a +\lambda_a{}^bM_b\,,
\end{eqnarray}
so that we must have the off-shell Ward identities 
\begin{eqnarray}
\hat e_\chi{}_\mu^a T^\mu & = & \tau_\nu e^{\mu a}T_\chi{}^\nu{}_\mu\,,\\
0 & = & \hat e_\chi{}^{[a}_\nu e^{b]\mu}T_\chi{}^\nu{}_\mu\,,\\
\langle O_\chi\rangle & = & \frac{1}{e}\partial_\mu\left(eT^\mu\right)\,.
\end{eqnarray}
Hence we obtain the following relation between the 5- and 4-dimensional vevs
\begin{eqnarray}
\frac{1}{2}t_{AB}\delta\gamma_{(0)}^{AB}+\langle O_\psi\rangle\delta\psi_{(0)} & = & -\tau_\nu T_\chi{}^\nu{}_\mu\delta\hat v_\chi^\mu-\left(\tau_{(\mu}\hat h^\chi_{\nu)\rho}\hat v_\chi^\sigma T_\chi{}^\rho{}_\sigma\right)\delta h^{\mu\nu}\nonumber\\
&&+\frac{1}{2}\left(\hat h^\chi_{\mu\rho}\hat h^\chi_{\nu\lambda}h^{\lambda\sigma} T_\chi{}^\rho{}_\sigma\right)\delta h^{\mu\nu}+\tau_\mu T^\mu\delta\tilde\Phi_\chi\,.
\end{eqnarray}
Using the same reasoning as in section \ref{subsec:nullredEMT} we conclude from this that the relation between the 5- and 4-dimensional vevs can be summarized as
\begin{eqnarray}
t^{\mu u} & = & 2\tilde\Phi T^\mu-\hat v^\nu\left(T_\chi{}^\mu{}_\nu-T^\mu\partial_\nu\chi\right)\,,\\
t^{\mu\nu} & = & -\hat v^\mu T^\nu+h^{\mu\rho}\left(T_\chi{}^\nu{}_\rho-T^\nu\partial_\rho\chi\right)\,.
\end{eqnarray}
Note that $T_\chi{}^\mu{}_\nu-T^\mu\partial_\nu\chi$ is independent of $\chi$ because we absorbed $T^\mu\partial_\nu\chi$ into the definition of $T_\chi{}^\mu{}_\nu$ (see also \eqref{eq:chiT}).
Put another way we can use equations \eqref{eq:nullredt1}--\eqref{eq:nullredt7} with 
\begin{equation}\label{eq:TandTchi}
T^\mu{}_\nu=T_\chi{}^\mu{}_\nu-T^\mu\partial_\nu\chi\,. 
\end{equation}
The Ward identities are then obtained by the dimensional reduction of \eqref{eq:tracet} and \eqref{eq:divt} using \eqref{eq:5d4dsource} and equations \eqref{eq:nullredt1}--\eqref{eq:nullredt7} with $T^\mu{}_\nu=T_\chi{}^\mu{}_\nu-T^\mu\partial_\nu\chi$. On a flat boundary with $\tau_\mu=\delta_\mu^t$, $h_{\mu\nu}=\delta_{ij}\delta^i_\mu\delta^j_\nu$, $m_\mu=0$ and $\chi=0$ this becomes
\begin{eqnarray}
2T^t{}_t+T^i{}_i & = & 0\,,\label{eq:WILifhydro1}\\
\partial_\mu T^\mu{}_\nu & = & 0\,,\label{eq:WILifhydro2}\\
\partial_\mu T^\mu & = & \langle O_\chi\rangle\,.\label{eq:WILifhydro3}
\end{eqnarray}

\subsection{Consistency of the reduction}\label{subsec:consistency}

In this subsection we will show that the Scherk--Schwarz reduction \eqref{eq:KK} and \eqref{eq:KK2} is consistent. We performed the reduction at the level of the action in section \ref{subsec:dimredTNC}. It remains to show that also the equations of motion of the 5-dimensional action reduce correctly. The 5-dimensional equations of motion \eqref{eq:Einsteineqs}and \eqref{eq:phieom} can be written as
\begin{eqnarray}
R^{(5)}_{\mathcal{M}\mathcal{N}} & = & -4\mathcal{G}_{\mathcal{M}\mathcal{N}}+\frac{1}{2}\partial_{\mathcal{M}}\psi\partial_{\mathcal{N}}\psi\,,\\
0 & = & \partial_{\mathcal{M}}\left(\sqrt{-\mathcal{G}}\mathcal{G}^{\mathcal{M}\mathcal{N}}\partial_{\mathcal{N}}\psi\right) \,,
\end{eqnarray}
where the superscript on the Ricci tensor is used to distinguish its $MN$ component from the 4-dimensional Ricci tensor $R^{(4)}_{MN}$.

The Kaluza--Klein ansatz for the metric \eqref{eq:KK} tells us that
\begin{eqnarray}
&&\mathcal{G}_{MN}=e^{-\Phi}g_{MN}+e^{2\Phi}A_M A_N\,,\qquad\mathcal{G}_{Mu}=e^{2\Phi}A_M\,,\qquad \mathcal{G}_{uu}=e^{2\Phi}\,,\\
&&\mathcal{G}^{MN}=e^{\Phi}g^{MN}\,,\qquad g^{Mu}=-e^\Phi A^M\,,\qquad\mathcal{G}^{uu}=e^{-2\Phi}+e^\Phi A^MA_M\,.
\end{eqnarray} 
Further we have $\sqrt{-\mathcal{G}}=e^{-\Phi}\sqrt{-g}$. The reduction of the 5-dimensional Ricci tensor follows from standard results on circle reductions of gravity (see for example \cite{Ortin:2004ms}). The components of the 5-dimensional Ricci tensor can be written as follows
\begin{eqnarray}
R^{(5)}_{uu} & = & -e^{3\Phi}\square\Phi+\frac{1}{4} e^{6\Phi}F^2\,,\label{eq:uu1}\\
R^{(5)}_{uM} & = & R^{(5)}_{uu}A_M+\frac{1}{2}\nabla^N\left(e^{3\Phi}F_{MN}\right)\,,\label{eq:uM1}\\
R^{(5)}_{MN} & = & A_M R^{(5)}_{uN}+A_N R^{(5)}_{uM}-A_MA_N R^{(5)}_{uu}+R^{(4)}_{MN}-\frac{3}{2}\partial_M\Phi\partial_N\Phi\nonumber\\
&&+\frac{1}{2}g_{MN}\square\Phi-\frac{1}{2}e^{3\Phi}F_{MP}F_{N}{}^P\,.\label{eq:uMN1}
\end{eqnarray}
Using the Scherk--Schwarz reduction ansatz for $\psi$ given in \eqref{eq:KK2} we also have
\begin{eqnarray}
R^{(5)}_{uu} & = & -4e^{2\Phi}+2\,,\label{eq:uu2}\\
R^{(5)}_{uM} & = & -4e^{2\Phi}A_M+2\partial_M\Xi\,,\label{eq:uM2}\\
R^{(5)}_{MN} & = & -4e^{-\Phi}g_{MN}-4e^{2\Phi}A_M A_N+2\partial_M\Xi\partial_N\Xi\,.\label{eq:uMN2}
\end{eqnarray}

It is now straightforward to verify that combining \eqref{eq:uu1} and \eqref{eq:uu2} leads to the equation of motion for $\Phi$ given in \eqref{eq:eomupliftable2}. Continuing with \eqref{eq:uM1} and \eqref{eq:uM2} we obtain the equation of motion for $B_M=A_M-\partial_M\Xi$ given in \eqref{eq:eomupliftable2}. Finally the equations \eqref{eq:uMN1} and \eqref{eq:uMN1} lead to the trace-reversed versions of the Einstein equation given in \eqref{eq:eomupliftable1}. We also have the 5-dimensional equation of motion for $\psi$. This can be seen to reduce to $\partial_M\left(\sqrt{-g}B^M\right)=0$ which is a consequence of \eqref{eq:eomupliftable2}. We have hereby shown that the reduction \eqref{eq:KK} and \eqref{eq:KK2} is consistent.

\subsection{Comparison to other approaches}\label{subsec:comparison}

The works \cite{Chemissany:2014xpa,Chemissany:2014xsa} and \cite{Hartong:2014oma,Hartong:2015wxa} both study asymptotically locally Lifshitz solutions of the EPD model. The setup of  \cite{Chemissany:2014xpa,Chemissany:2014xsa} includes in principle what we refer to as the upliftable model but does not study it explicitly. Below we will make a first attempt at a comparison between the two approaches. We will do this for the general class of EPD models for as much as possible. Some statements will however be more specific for the case of the upliftable model. 

In the notation of \cite{Chemissany:2014xpa,Chemissany:2014xsa} the solution to the equations of motion of the EPD model near a Lifshitz boundary is written as
\begin{eqnarray}
ds^2 & = & dr^2+\gamma_{ij}dx^i dx^j\,,\\
A & = & A_i dx^i\,,\qquad B=A-d\omega\,.
\end{eqnarray}
The $U(1)$ gauge transformations have been partially fixed by setting $A_r=0$. In \cite{Hartong:2014oma,Hartong:2015wxa} we make the same gauge choice only for $z=2$. In our notation we would replace the $i$ index by a $\mu$ index and replace $r$ by $\log r$. Further $\omega$ here is denoted by $\Xi${}\footnote{The source $\omega_{(0)}$ of the St\"uckelberg field $\omega$ is what we call $\chi$.} and $\gamma_{ij}$ is called $h_{\mu\nu}$ here. 

In \cite{Chemissany:2014xpa,Chemissany:2014xsa}  a radial gauge ($F_2=1$) is employed for the metric while in \cite{Hartong:2014oma,Hartong:2015wxa} we allow for a general function in the $rr$ component of the metric. For example for the upliftable model it is more natural to work with a gauge in which{}\footnote{Similar gauge choices for $F_2$ are also important for some other EPD models that do not admit an uplift (see the $z=2$ and $\Delta=0$ cases discussed in \cite{Hartong:2014oma}).}  $F_2=e^{-\Phi}$ so that the 5-dimensional uplifted asymptotically locally AdS metric is written in radial gauge. This difference is more than just a matter of choice because we have shown in \cite{Christensen:2013rfa} that one cannot transform to the $F_2=1$ gauge unless the leading term in the expansion of $\Phi$, that we call $\phi$, vanishes\footnote{For general EPD models we set $\Phi\simeq r^\Delta\phi$ \cite{Hartong:2014oma,Hartong:2015wxa} where the value of $\Delta$ depends on the details of the EPD action. To the best of our knowledge this $\Delta$ parameter does not appear explicitly in \cite{Chemissany:2014xpa,Chemissany:2014xsa}. However there is a comment below their equation (5.25) stating that the asymptotic form of the dilaton depends on the potential which is essentially allowing for a $\Delta$ in the fall-off of the dilaton.}. This is not always the case and when we impose this extra condition it leads via \eqref{eq:constraint} to the condition that $\tau_\mu$ is hypersurface orthogonal. Hence if we make the assumption that the asymptotically locally Lifshitz boundary conditions of \cite{Christensen:2013rfa} are compatible with radial gauge we need to put the source $\phi=0$. 

In \cite{Chemissany:2014xpa,Chemissany:2014xsa} the metric $\gamma_{ij}$ is written in the ADM decomposition as
\begin{equation}
\gamma_{ij}dx^i dx^j=-n^2dt^2+\sigma_{ab}\left(dx^a+n^adt\right)\left(dx^b+n^bdt\right)\,,
\end{equation}
where $a$ labels the number of spatial dimensions which here is $d=2$. In order to make contact with the way we set up the definition of the sources we write the ADM decomposition in terms of vielbeins as follows
\begin{equation}
\gamma_{ij}dx^idx^j=-E^0E^0+\delta_{\underline{a}\underline{b}}E^{\underline{a}}E^{\underline{b}}\,,
\end{equation}
where underlined indices $\underline{a}$ refer to flat tangent space indices that take as many values as there are spatial coordinates. We keep here with the notation of \cite{Chemissany:2014xpa,Chemissany:2014xsa}. Since these works do not use tangent space indices we introduced these underlined indices only in this section for the sake of comparison. We can take without loss of generality
\begin{equation}
E^0=ndt\,,\qquad E^{\underline{a}}=e^{\underline{a}}_a\left(dx^a+n^a dt\right)\,.
\end{equation}
This allows us to establish the following dictionary between the sources in \cite{Chemissany:2014xpa,Chemissany:2014xsa} and those defined in \cite{Hartong:2014oma,Hartong:2015wxa}
\begin{eqnarray}
&&\tau_{\mu}=\left(n_{(0)}\,,0\right)\,,\qquad v^\mu=n_{(0)}^{-1}\left(-1\,, n_{(0)}^a\right)\,,\\
&&h_{\mu\nu}dx^\mu dx^\nu=g_{(0)ab}\left(dx^a+n_{(0)}^adt\right)\left(dx^b+n_{(0)}^bdt\right)\,.
\end{eqnarray}
Hence in \cite{Chemissany:2014xpa,Chemissany:2014xsa} the source $\tau_\mu$ is always taken to be hypersurface orthogonal. 

In the work \cite{Hartong:2014oma,Hartong:2015wxa} we also introduce the Newton--Cartan vector $m_\mu$ as a source or rather the $U(1)$ invariant combination $M_\mu=m_\mu-\partial_\mu\chi$. By fixing local tangent space transformations (with parameter $\lambda_a$) we can fix all but one component of $M_\mu$. The remaining component is related to $\tilde\Phi_\chi$ which is defined by \eqref{eq:tildePhi} with $m_\mu$ replaced by $M_\mu$. The scalar $\tilde\Phi$ has scaling weight $2(z-1)$. Similarly there is a scalar source in the work of \cite{Chemissany:2014xpa,Chemissany:2014xsa} that is denoted by $\psi$ following \cite{Ross:2011gu}. The main difference between the approach of \cite{Chemissany:2014xpa,Chemissany:2014xsa} and \cite{Hartong:2014oma,Hartong:2015wxa} lies in the fact that the dilatation weight of $\psi$ denoted by $\Delta_-$ does not in general agree with the dilatation weight of $\tilde\Phi$. The number of sources (when comparing both approaches in radial gauge and taking $\tau_\mu$ to be hypersurface orthogonal) thus agrees but for one of them the scaling dimensions differ. 

To see where $\tilde\Phi$ appears in our near-boundary expansion we consider purely radial solutions, like we studied in section \ref{subsec:asympsol}. Recall that in section \ref{subsec:asympsol} we set $\tilde\Phi=0$ by hand. If we do not do this then we obtain, using the results of appendices \ref{subsec:nullred} and \ref{sec:FGexpansions},
\begin{eqnarray}
\gamma_{(2)AB} & = & \delta^u_A\delta^u_B-\frac{1}{3}\tilde\Phi\gamma_{(0)AB}\,,\\
\gamma_{(4,1)AB} & = & \frac{4}{3}\tilde\Phi\delta^u_A\delta^u_B-\frac{2}{3}\tilde\Phi^2\gamma_{(0)AB}\,,\\
\gamma_{(4)AB} & = & -\frac{1}{2}\tilde\Phi\delta^u_A\delta^u_B+\frac{5}{18}\tilde\Phi^2\gamma_{(0)AB}-\frac{1}{4}t_{AB}\,,\\
\psi_{(2)} & = & \psi_{(4,1)} = \psi_{(4)} = 0\,.
\end{eqnarray}
Components such as $\gamma_{(4,1)AB}$ correspond to logarithmic terms in the expansion. This implies that for example the expansions of the matter fields become
\begin{eqnarray}
\Phi & = & \frac{2}{3}\tilde\Phi r^2\log r-\frac{1}{8}(\rho+2\tilde\Phi)r^2+\ldots\,,\\
A_\mu & = & r^{-2}\tau_\mu-\frac{4}{3}\log r\tilde\Phi\tau_\mu+\frac{1}{6}\tilde\Phi\tau_\mu+\frac{1}{4}\rho\tau_\mu+\ldots\,,
\end{eqnarray}
where the dots denote subleading terms. 

It would be nice to make a direct comparison between the case studied here where $V$ is a sum of two exponential potentials as given in \eqref{eq:upliftablemodel}. However the case where $V$ is the sum of two exponentials is not explicitly studied in \cite{Chemissany:2014xpa,Chemissany:2014xsa} so this would have to be worked out first{}\footnote{There exists another model used in \cite{Christensen:2013rfa} that contains two dilatons that can be obtained by a similar Scherk--Schwarz reduction as used in the present work admitting a $z=2$ Lifshitz solution. This model is related in \cite{Chemissany:2012du} (see around equation (6.40)) to an EPD model with a single exponential potential, which is a case explicitly worked out in that paper. This is done by setting a linear combination of these two scalars equal to a constant, which, however,  is not a consistent truncation of the model discussed in \cite{Chemissany:2012du}. The relation (6.40) of \cite{Chemissany:2012du} only holds asymptotically at leading order in a near-boundary expansion.}. 

\addcontentsline{toc}{section}{References}

\bibliography{Lifshitzhydro2}

\providecommand{\href}[2]{#2}\begingroup\raggedright\begin{thebibliography}{10}

\bibitem{Policastro:2001yc}
G.~Policastro, D.~T. Son, and A.~O. Starinets, ``{The Shear viscosity of
  strongly coupled N=4 supersymmetric Yang-Mills plasma},''
  \href{http://dx.doi.org/10.1103/PhysRevLett.87.081601}{{\em Phys.Rev.Lett.}
  {\bf 87} (2001)  081601},
\href{http://arxiv.org/abs/hep-th/0104066}{{\tt arXiv:hep-th/0104066
  [hep-th]}}.

\bibitem{Kovtun:2004de}
P.~Kovtun, D.~T. Son, and A.~O. Starinets, ``{Viscosity in strongly interacting
  quantum field theories from black hole physics},''
  \href{http://dx.doi.org/10.1103/PhysRevLett.94.111601}{{\em Phys. Rev. Lett.}
  {\bf 94} (2005)  111601},
\href{http://arxiv.org/abs/hep-th/0405231}{{\tt arXiv:hep-th/0405231
  [hep-th]}}.

\bibitem{Son:2007vk}
D.~T. Son and A.~O. Starinets, ``{Viscosity, Black Holes, and Quantum Field
  Theory},''
  \href{http://dx.doi.org/10.1146/annurev.nucl.57.090506.123120}{{\em Ann. Rev.
  Nucl. Part. Sci.} {\bf 57} (2007)  95--118},
\href{http://arxiv.org/abs/0704.0240}{{\tt arXiv:0704.0240 [hep-th]}}.

\bibitem{Bhattacharyya:2008jc}
S.~Bhattacharyya, V.~E. Hubeny, S.~Minwalla, and M.~Rangamani, ``{Nonlinear
  Fluid Dynamics from Gravity},''
  \href{http://dx.doi.org/10.1088/1126-6708/2008/02/045}{{\em JHEP} {\bf 0802}
  (2008)  045},
\href{http://arxiv.org/abs/0712.2456}{{\tt arXiv:0712.2456 [hep-th]}}.

\bibitem{Baier:2007ix}
R.~Baier, P.~Romatschke, D.~T. Son, A.~O. Starinets, and M.~A. Stephanov,
  ``{Relativistic viscous hydrodynamics, conformal invariance, and
  holography},'' \href{http://dx.doi.org/10.1088/1126-6708/2008/04/100}{{\em
  JHEP} {\bf 04} (2008)  100},
\href{http://arxiv.org/abs/0712.2451}{{\tt arXiv:0712.2451 [hep-th]}}.

\bibitem{Rangamani:2009xk}
M.~Rangamani, ``{Gravity and Hydrodynamics: Lectures on the fluid-gravity
  correspondence},''
  \href{http://dx.doi.org/10.1088/0264-9381/26/22/224003}{{\em Class. Quant.
  Grav.} {\bf 26} (2009)  224003},
\href{http://arxiv.org/abs/0905.4352}{{\tt arXiv:0905.4352 [hep-th]}}.

\bibitem{Emparan:2009cs}
R.~Emparan, T.~Harmark, V.~Niarchos, and N.~A. Obers, ``{World-Volume Effective
  Theory for Higher-Dimensional Black Holes},''
  \href{http://dx.doi.org/10.1103/PhysRevLett.102.191301}{{\em Phys. Rev.
  Lett.} {\bf 102} (2009)  191301},
\href{http://arxiv.org/abs/0902.0427}{{\tt arXiv:0902.0427 [hep-th]}}.

\bibitem{Emparan:2009at}
R.~Emparan, T.~Harmark, V.~Niarchos, and N.~A. Obers, ``{Essentials of
  Blackfold Dynamics},'' \href{http://dx.doi.org/10.1007/JHEP03(2010)063}{{\em
  JHEP} {\bf 03} (2010)  063},
\href{http://arxiv.org/abs/0910.1601}{{\tt arXiv:0910.1601 [hep-th]}}.

\bibitem{Camps:2010br}
J.~Camps, R.~Emparan, and N.~Haddad, ``{Black Brane Viscosity and the
  Gregory-Laflamme Instability},''
  \href{http://dx.doi.org/10.1007/JHEP05(2010)042}{{\em JHEP} {\bf 05} (2010)
  042},
\href{http://arxiv.org/abs/1003.3636}{{\tt arXiv:1003.3636 [hep-th]}}.

\bibitem{Emparan:2013ila}
R.~Emparan, V.~E. Hubeny, and M.~Rangamani, ``{Effective hydrodynamics of black
  D3-branes},'' \href{http://dx.doi.org/10.1007/JHEP06(2013)035}{{\em JHEP}
  {\bf 06} (2013)  035},
\href{http://arxiv.org/abs/1303.3563}{{\tt arXiv:1303.3563 [hep-th]}}.

\bibitem{Son:2008ye}
D.~Son, ``{Toward an AdS/cold atoms correspondence: A Geometric realization of
  the Schrodinger symmetry},''
  \href{http://dx.doi.org/10.1103/PhysRevD.78.046003}{{\em Phys.Rev.} {\bf D78}
  (2008)  046003},
\href{http://arxiv.org/abs/0804.3972}{{\tt arXiv:0804.3972 [hep-th]}}.

\bibitem{Balasubramanian:2008dm}
K.~Balasubramanian and J.~McGreevy, ``{Gravity duals for non-relativistic
  CFTs},'' \href{http://dx.doi.org/10.1103/PhysRevLett.101.061601}{{\em
  Phys.Rev.Lett.} {\bf 101} (2008)  061601},
\href{http://arxiv.org/abs/0804.4053}{{\tt arXiv:0804.4053 [hep-th]}}.

\bibitem{Kachru:2008yh}
S.~Kachru, X.~Liu, and M.~Mulligan, ``{Gravity Duals of Lifshitz-like Fixed
  Points},'' \href{http://dx.doi.org/10.1103/PhysRevD.78.106005}{{\em
  Phys.Rev.} {\bf D78} (2008)  106005},
\href{http://arxiv.org/abs/0808.1725}{{\tt arXiv:0808.1725 [hep-th]}}.

\bibitem{Taylor:2008tg}
M.~Taylor, ``{Non-relativistic holography},''
\href{http://arxiv.org/abs/0812.0530}{{\tt arXiv:0812.0530 [hep-th]}}.

\bibitem{Taylor:2015glc}
M.~Taylor, ``{Lifshitz holography},'' {\em Class. Quant. Grav.} {\bf 33} (2016)
  no.~3, 033001,
\href{http://arxiv.org/abs/1512.03554}{{\tt arXiv:1512.03554 [hep-th]}}.

\bibitem{Herzog:2008wg}
C.~P. Herzog, M.~Rangamani, and S.~F. Ross, ``{Heating up Galilean
  holography},'' \href{http://dx.doi.org/10.1088/1126-6708/2008/11/080}{{\em
  JHEP} {\bf 11} (2008)  080},
\href{http://arxiv.org/abs/0807.1099}{{\tt arXiv:0807.1099 [hep-th]}}.

\bibitem{Maldacena:2008wh}
J.~Maldacena, D.~Martelli, and Y.~Tachikawa, ``{Comments on string theory
  backgrounds with non-relativistic conformal symmetry},''
  \href{http://dx.doi.org/10.1088/1126-6708/2008/10/072}{{\em JHEP} {\bf 0810}
  (2008)  072},
\href{http://arxiv.org/abs/0807.1100}{{\tt arXiv:0807.1100 [hep-th]}}.

\bibitem{Adams:2008wt}
A.~Adams, K.~Balasubramanian, and J.~McGreevy, ``{Hot Spacetimes for Cold
  Atoms},'' \href{http://dx.doi.org/10.1088/1126-6708/2008/11/059}{{\em JHEP}
  {\bf 11} (2008)  059},
\href{http://arxiv.org/abs/0807.1111}{{\tt arXiv:0807.1111 [hep-th]}}.

\bibitem{Rangamani:2008gi}
M.~Rangamani, S.~F. Ross, D.~T. Son, and E.~G. Thompson, ``{Conformal
  non-relativistic hydrodynamics from gravity},''
  \href{http://dx.doi.org/10.1088/1126-6708/2009/01/075}{{\em JHEP} {\bf 01}
  (2009)  075},
\href{http://arxiv.org/abs/0811.2049}{{\tt arXiv:0811.2049 [hep-th]}}.

\bibitem{Hoyos:2013eza}
C.~Hoyos, B.~S. Kim, and Y.~Oz, ``{Lifshitz Hydrodynamics},''
  \href{http://dx.doi.org/10.1007/JHEP11(2013)145}{{\em JHEP} {\bf 1311} (2013)
   145},
\href{http://arxiv.org/abs/1304.7481}{{\tt arXiv:1304.7481 [hep-th]}}.

\bibitem{Hoyos:2013qna}
C.~Hoyos, B.~S. Kim, and Y.~Oz, ``{Lifshitz Field Theories at Non-Zero
  Temperature, Hydrodynamics and Gravity},''
  \href{http://dx.doi.org/10.1007/JHEP03(2014)029}{{\em JHEP} {\bf 1403} (2014)
   029},
\href{http://arxiv.org/abs/1309.6794}{{\tt arXiv:1309.6794 [hep-th]}}.

\bibitem{Hoyos:2013cba}
C.~Hoyos, B.~S. Kim, and Y.~Oz, ``{Bulk Viscosity in Holographic Lifshitz
  Hydrodynamics},'' \href{http://dx.doi.org/10.1007/JHEP03(2014)050}{{\em JHEP}
  {\bf 03} (2014)  050},
\href{http://arxiv.org/abs/1312.6380}{{\tt arXiv:1312.6380 [hep-th]}}.

\bibitem{Kiritsis:2015doa}
E.~Kiritsis and Y.~Matsuo, ``{Charge-hyperscaling violating Lifshitz
  hydrodynamics from black-holes},''
  \href{http://dx.doi.org/10.1007/JHEP12(2015)076}{{\em JHEP} {\bf 12} (2015)
  076},
\href{http://arxiv.org/abs/1508.02494}{{\tt arXiv:1508.02494 [hep-th]}}.

\bibitem{Christensen:2013lma}
M.~H. Christensen, J.~Hartong, N.~A. Obers, and B.~Rollier, ``{Torsional
  Newton-Cartan Geometry and Lifshitz Holography},''
  \href{http://dx.doi.org/10.1103/PhysRevD.89.061901}{{\em Phys.Rev.} {\bf D89}
  (2014)  061901},
\href{http://arxiv.org/abs/1311.4794}{{\tt arXiv:1311.4794 [hep-th]}}.

\bibitem{Christensen:2013rfa}
M.~H. Christensen, J.~Hartong, N.~A. Obers, and B.~Rollier, ``{Boundary
  Stress-Energy Tensor and Newton-Cartan Geometry in Lifshitz Holography},''
  \href{http://dx.doi.org/10.1007/JHEP01(2014)057}{{\em JHEP} {\bf 1401} (2014)
   057},
\href{http://arxiv.org/abs/1311.6471}{{\tt arXiv:1311.6471 [hep-th]}}.

\bibitem{Hartong:2014oma}
J.~Hartong, E.~Kiritsis, and N.~A. Obers, ``{Lifshitz spaceÐtimes for
  Schr\"odinger holography},''
  \href{http://dx.doi.org/10.1016/j.physletb.2015.05.010}{{\em Phys. Lett.}
  {\bf B746} (2015)  318--324},
\href{http://arxiv.org/abs/1409.1519}{{\tt arXiv:1409.1519 [hep-th]}}.

\bibitem{Hartong}
J.~Hartong, N.~Obers, W.~Sybesma, and S.~Vandoren, ``{to appear},''.

\bibitem{Hartong:2015wxa}
J.~Hartong, E.~Kiritsis, and N.~A. Obers, ``{Field Theory on Newton-Cartan
  Backgrounds and Symmetries of the Lifshitz Vacuum},''
  \href{http://dx.doi.org/10.1007/JHEP08(2015)006}{{\em JHEP} {\bf 08} (2015)
  006},
\href{http://arxiv.org/abs/1502.00228}{{\tt arXiv:1502.00228 [hep-th]}}.

\bibitem{Gath:2012pg}
J.~Gath, J.~Hartong, R.~Monteiro, and N.~A. Obers, ``{Holographic Models for
  Theories with Hyperscaling Violation},''
  \href{http://dx.doi.org/10.1007/JHEP04(2013)159}{{\em JHEP} {\bf 1304} (2013)
   159},
\href{http://arxiv.org/abs/1212.3263}{{\tt arXiv:1212.3263 [hep-th]}}.

\bibitem{Gouteraux:2012yr}
B.~Gouteraux and E.~Kiritsis, ``{Quantum critical lines in holographic phases
  with (un)broken symmetry},''
  \href{http://dx.doi.org/10.1007/JHEP04(2013)053}{{\em JHEP} {\bf 1304} (2013)
   053},
\href{http://arxiv.org/abs/1212.2625}{{\tt arXiv:1212.2625 [hep-th]}}.

\bibitem{Charmousis:2010zz}
C.~Charmousis, B.~Gouteraux, B.~Kim, E.~Kiritsis, and R.~Meyer, ``{Effective
  Holographic Theories for low-temperature condensed matter systems},''
  \href{http://dx.doi.org/10.1007/JHEP11(2010)151}{{\em JHEP} {\bf 1011} (2010)
   151},
\href{http://arxiv.org/abs/1005.4690}{{\tt arXiv:1005.4690 [hep-th]}}.

\bibitem{Bertoldi:2009vn}
G.~Bertoldi, B.~A. Burrington, and A.~Peet, ``{Black Holes in asymptotically
  Lifshitz spacetimes with arbitrary critical exponent},''
  \href{http://dx.doi.org/10.1103/PhysRevD.80.126003}{{\em Phys.Rev.} {\bf D80}
  (2009)  126003},
\href{http://arxiv.org/abs/0905.3183}{{\tt arXiv:0905.3183 [hep-th]}}.

\bibitem{Bertoldi:2009dt}
G.~Bertoldi, B.~A. Burrington, and A.~W. Peet, ``{Thermodynamics of black
  branes in asymptotically Lifshitz spacetimes},''
  \href{http://dx.doi.org/10.1103/PhysRevD.80.126004}{{\em Phys.Rev.} {\bf D80}
  (2009)  126004},
\href{http://arxiv.org/abs/0907.4755}{{\tt arXiv:0907.4755 [hep-th]}}.

\bibitem{Hartong:2014pma}
J.~Hartong, E.~Kiritsis, and N.~A. Obers, ``{Schr\"odinger Invariance from
  Lifshitz Isometries in Holography and Field Theory},''
  \href{http://dx.doi.org/10.1103/PhysRevD.92.066003}{{\em Phys. Rev.} {\bf
  D92} (2015)  066003},
\href{http://arxiv.org/abs/1409.1522}{{\tt arXiv:1409.1522 [hep-th]}}.

\bibitem{Chemissany:2012du}
W.~Chemissany, D.~Geissbuhler, J.~Hartong, and B.~Rollier, ``{Holographic
  Renormalization for z=2 Lifshitz Space-Times from AdS},''
  \href{http://dx.doi.org/10.1088/0264-9381/29/23/235017}{{\em
  Class.Quant.Grav.} {\bf 29} (2012)  235017},
\href{http://arxiv.org/abs/1205.5777}{{\tt arXiv:1205.5777 [hep-th]}}.

\bibitem{Eisenhart}
L.~P. Eisenhart, ``{Dynamical Trajectories and Geodesics},'' {\em Annals of
  Mathematics} {\bf 30} (1928)  591.

\bibitem{Kuenzle:1972zw}
H.~Kuenzle, ``{Galilei and lorentz structures on space-time - comparison of the
  corresponding geometry and physics},''
{\em Annales Poincare Phys.Theor.} {\bf 17} (1972)  337--362.

\bibitem{Julia:1994bs}
B.~Julia and H.~Nicolai, ``{Null Killing vector dimensional reduction and
  Galilean geometrodynamics},''
  \href{http://dx.doi.org/10.1016/0550-3213(94)00584-2}{{\em Nucl.Phys.} {\bf
  B439} (1995)  291--326},
\href{http://arxiv.org/abs/hep-th/9412002}{{\tt arXiv:hep-th/9412002
  [hep-th]}}.

\bibitem{Bergshoeff:2014uea}
E.~A. Bergshoeff, J.~Hartong, and J.~Rosseel, ``{Torsional Newton--Cartan
  geometry and the Schr\"odinger algebra},''
  \href{http://dx.doi.org/10.1088/0264-9381/32/13/135017}{{\em Class. Quant.
  Grav.} {\bf 32} (2015) no.~13, 135017},
\href{http://arxiv.org/abs/1409.5555}{{\tt arXiv:1409.5555 [hep-th]}}.

\bibitem{Hartong:2010ec}
J.~Hartong and B.~Rollier, ``{Asymptotically Schroedinger Space-Times: TsT
  Transformations and Thermodynamics},''
  \href{http://dx.doi.org/10.1007/JHEP01(2011)084}{{\em JHEP} {\bf 1101} (2011)
   084},
\href{http://arxiv.org/abs/1009.4997}{{\tt arXiv:1009.4997 [hep-th]}}.

\bibitem{Papadimitriou:2005ii}
I.~Papadimitriou and K.~Skenderis, ``{Thermodynamics of asymptotically locally
  AdS spacetimes},''
  \href{http://dx.doi.org/10.1088/1126-6708/2005/08/004}{{\em JHEP} {\bf 0508}
  (2005)  004},
\href{http://arxiv.org/abs/hep-th/0505190}{{\tt arXiv:hep-th/0505190
  [hep-th]}}.

\bibitem{Armas:2015nea}
J.~Armas and M.~Blau, ``{New Geometries for Black Hole Horizons},''
  \href{http://dx.doi.org/10.1007/JHEP07(2015)048}{{\em JHEP} {\bf 07} (2015)
  048},
\href{http://arxiv.org/abs/1504.01393}{{\tt arXiv:1504.01393 [hep-th]}}.

\bibitem{Chemissany:2014xpa}
W.~Chemissany and I.~Papadimitriou, ``{Generalized dilatation operator method
  for non-relativistic holography},''
  \href{http://dx.doi.org/10.1016/j.physletb.2014.08.057}{{\em Phys.Lett.} {\bf
  B737} (2014)  272--276},
\href{http://arxiv.org/abs/1405.3965}{{\tt arXiv:1405.3965 [hep-th]}}.

\bibitem{Chemissany:2014xsa}
W.~Chemissany and I.~Papadimitriou, ``{Lifshitz holography: The whole
  shebang},'' \href{http://dx.doi.org/10.1007/JHEP01(2015)052}{{\em JHEP} {\bf
  01} (2015)  052},
\href{http://arxiv.org/abs/1408.0795}{{\tt arXiv:1408.0795 [hep-th]}}.

\bibitem{Ross:2011gu}
S.~F. Ross, ``{Holography for asymptotically locally Lifshitz spacetimes},''
  \href{http://dx.doi.org/10.1088/0264-9381/28/21/215019}{{\em
  Class.Quant.Grav.} {\bf 28} (2011)  215019},
\href{http://arxiv.org/abs/1107.4451}{{\tt arXiv:1107.4451 [hep-th]}}.

\bibitem{Baggio:2011cp}
M.~Baggio, J.~de~Boer, and K.~Holsheimer, ``{Hamilton-Jacobi Renormalization
  for Lifshitz Spacetime},''
  \href{http://dx.doi.org/10.1007/JHEP01(2012)058}{{\em JHEP} {\bf 1201} (2012)
   058},
\href{http://arxiv.org/abs/1107.5562}{{\tt arXiv:1107.5562 [hep-th]}}.

\bibitem{Tarrio:2011de}
J.~Tarrio and S.~Vandoren, ``{Black holes and black branes in Lifshitz
  spacetimes},'' \href{http://dx.doi.org/10.1007/JHEP09(2011)017}{{\em JHEP}
  {\bf 1109} (2011)  017},
\href{http://arxiv.org/abs/1105.6335}{{\tt arXiv:1105.6335 [hep-th]}}.

\bibitem{Festuccia2016b}
G.~Festuccia, D.~Hansen, J.~Hartong, and N.~A. Obers, ``{Symmetries and
  Couplings of Non-Relativistic Electrodynamics (to appear)},''.

\bibitem{Griffin:2012qx}
T.~Griffin, P.~Horava, and C.~M. Melby-Thompson, ``{Lifshitz Gravity for
  Lifshitz Holography},''
  \href{http://dx.doi.org/10.1103/PhysRevLett.110.081602}{{\em Phys.Rev.Lett.}
  {\bf 110} (2013) no.~8, 081602},
\href{http://arxiv.org/abs/1211.4872}{{\tt arXiv:1211.4872 [hep-th]}}.

\bibitem{Janiszewski:2012nf}
S.~Janiszewski and A.~Karch, ``{String Theory Embeddings of Nonrelativistic
  Field Theories and Their Holographic Horava Gravity Duals},''
  \href{http://dx.doi.org/10.1103/PhysRevLett.110.081601}{{\em Phys.Rev.Lett.}
  {\bf 110} (2013) no.~8, 081601},
\href{http://arxiv.org/abs/1211.0010}{{\tt arXiv:1211.0010 [hep-th]}}.

\bibitem{Eling:2014saa}
C.~Eling and Y.~Oz, ``{Horava-Lifshitz Black Hole Hydrodynamics},''
  \href{http://dx.doi.org/10.1007/JHEP11(2014)067}{{\em JHEP} {\bf 11} (2014)
  067},
\href{http://arxiv.org/abs/1408.0268}{{\tt arXiv:1408.0268 [hep-th]}}.

\bibitem{Davison:2016auk}
R.~A. Davison, S.~Grozdanov, S.~Janiszewski, and M.~Kaminski, ``{Momentum and
  charge transport in non-relativistic holographic fluids from Ho\v{r}ava
  gravity},''
\href{http://arxiv.org/abs/1606.06747}{{\tt arXiv:1606.06747 [hep-th]}}.

\bibitem{Hartong:2015zia}
J.~Hartong and N.~A. Obers, ``{Ho\v{r}ava-Lifshitz gravity from dynamical
  Newton-Cartan geometry},''
  \href{http://dx.doi.org/10.1007/JHEP07(2015)155}{{\em JHEP} {\bf 07} (2015)
  155},
\href{http://arxiv.org/abs/1504.07461}{{\tt arXiv:1504.07461 [hep-th]}}.

\bibitem{Hartong:2016yrf}
J.~Hartong, Y.~Lei, and N.~A. Obers, ``{Non-Relativistic Chern-Simons Theories
  and Three-Dimensional Horava-Lifshitz Gravity},''
\href{http://arxiv.org/abs/1604.08054}{{\tt arXiv:1604.08054 [hep-th]}}.

\bibitem{Andringa:2010it}
R.~Andringa, E.~Bergshoeff, S.~Panda, and M.~de~Roo, ``{Newtonian Gravity and
  the Bargmann Algebra},''
  \href{http://dx.doi.org/10.1088/0264-9381/28/10/105011}{{\em
  Class.Quant.Grav.} {\bf 28} (2011)  105011},
\href{http://arxiv.org/abs/1011.1145}{{\tt arXiv:1011.1145 [hep-th]}}.

\bibitem{Jensen:2014aia}
K.~Jensen, ``{On the coupling of Galilean-invariant field theories to curved
  spacetime},''
\href{http://arxiv.org/abs/1408.6855}{{\tt arXiv:1408.6855 [hep-th]}}.

\bibitem{Festuccia2016a}
G.~Festuccia, D.~Hansen, J.~Hartong, and N.~A. Obers, ``{Torsional
  Newton-Cartan from the Noether Procedure (to appear)},''.

\bibitem{Hartong:2015xda}
J.~Hartong, ``{Gauging the Carroll Algebra and Ultra-Relativistic Gravity},''
  \href{http://dx.doi.org/10.1007/JHEP08(2015)069}{{\em JHEP} {\bf 08} (2015)
  069},
\href{http://arxiv.org/abs/1505.05011}{{\tt arXiv:1505.05011 [hep-th]}}.

\bibitem{Banerjee:2015hra}
N.~Banerjee, S.~Dutta, and A.~Jain, ``{Null Fluids - A New Viewpoint of
  Galilean Fluids},'' \href{http://dx.doi.org/10.1103/PhysRevD.93.105020}{{\em
  Phys. Rev.} {\bf D93} (2016) no.~10, 105020},
\href{http://arxiv.org/abs/1509.04718}{{\tt arXiv:1509.04718 [hep-th]}}.

\bibitem{Jensen:2014ama}
K.~Jensen, ``{Aspects of hot Galilean field theory},''
\href{http://arxiv.org/abs/1411.7024}{{\tt arXiv:1411.7024 [hep-th]}}.

\bibitem{deHaro:2000xn}
S.~de~Haro, S.~N. Solodukhin, and K.~Skenderis, ``{Holographic reconstruction
  of space-time and renormalization in the AdS / CFT correspondence},''
  \href{http://dx.doi.org/10.1007/s002200100381}{{\em Commun.Math.Phys.} {\bf
  217} (2001)  595--622},
\href{http://arxiv.org/abs/hep-th/0002230}{{\tt arXiv:hep-th/0002230
  [hep-th]}}.

\bibitem{Papadimitriou:2011qb}
I.~Papadimitriou, ``{Holographic Renormalization of general dilaton-axion
  gravity},'' \href{http://dx.doi.org/10.1007/JHEP08(2011)119}{{\em JHEP} {\bf
  1108} (2011)  119},
\href{http://arxiv.org/abs/1106.4826}{{\tt arXiv:1106.4826 [hep-th]}}.

\bibitem{FeffermanGraham}
C.~Fefferman and C.~R. Graham, ``{Conformal Invariants},'' {\em Elie Cartan et
  les Math\'ematiques d'aujourd'hui (Asterisque)} {\bf 1103} (1985)  95.

\bibitem{Ortin:2004ms}
T.~Ortin, ``{Gravity and strings},''
{\em Cambridge Unversity, Cambridge University Press} (2004)  .

\end{thebibliography}\endgroup
\bibliographystyle{newutphys}
\end{document}